\documentclass[prb,aps,twocolumn]{revtex4}

\usepackage{amsmath}
\usepackage{mathrsfs}
\usepackage[dvipdfmx]{graphicx}
\usepackage{bm}
\usepackage{here}

\DeclareMathOperator{\trace}{Tr}

\begin{document}

\title{Theory of tunneling conductance of anomalous Rashba 
metal / superconductor junctions }

\author{Toshiyuki Fukumoto}
\affiliation{Department of Applied Physics, Nagoya University, Nagoya 464-8603, Japan}
\author{Katsuhisa Taguchi}
\affiliation{Department of Applied Physics, Nagoya University, Nagoya 464-8603, Japan}
\author{Shingo Kobayashi}
\affiliation{Department of Applied Physics, Nagoya University, Nagoya 464-8603, Japan}
\author{Yukio Tanaka}
\affiliation{Department of Applied Physics, Nagoya University, Nagoya 464-8603, Japan}

\begin{abstract}
We theoretically study the charge conductance in 
anomalous Rashba metal (ARM)/superconductor junctions for various types of the pairing symmetries in the superconductor. 
The exotic state dubbed ARM, where one of the spin resolved Fermi surface is absent, is realized
when the chemical potential is tuned both in the presence of Rashba spin-orbit interaction (RSOI) and an exchange field. 
Although a fully polarized ferromagnet metal (FPFM) is also a system 
where the electron's spin degrees of a freedom is reduced to be half, 
the electrons in an ARM have distinct features from those in FPFM. 
For the ARM/spin-singlet superconductor junctions, 
the obtained tunneling conductance within the bulk energy gap is enhanced 
with the increase in the magnitude of the RSOI. 
In particular, in ARM/$d_{xy}$-wave superconductor junctions, 
the zero bias conductance peak 
is enhanced owing to the presence of the RSOI. 
For ARM/$p_{x}$-wave superconductor junctions, 
the condition of the existence of the zero bias conductance peak is 
significantly sensitive to the direction of the $\bm{d}$-vector of the $p_{x}$-wave 
superconductor. 
Furthermore, the obtained conductance in ARM/chiral $p$-wave superconductor 
junctions shows different behaviors as compared to those in ARM/helical $p$-wave superconductor junctions. 
This feature gives a guide to determine the spin structure 
of the Cooper pair in spin-triplet superconductor Sr$_{2}$RuO$_{4}$.

\end{abstract}

\pacs{pacs}
\maketitle

\section{Introduction \label{Introduction}}
\begin{figure}[b]
	\centering 
	\includegraphics[width=7.5cm]{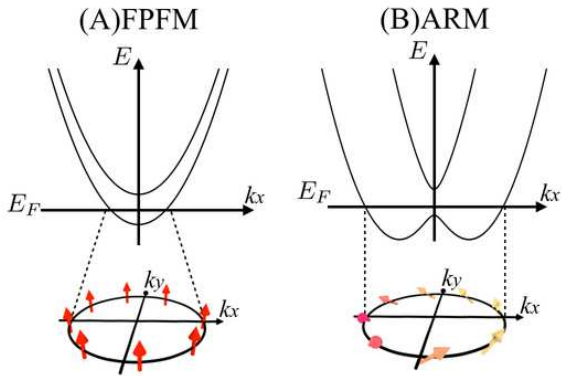}\\
	\caption{(Color online) Schematic illustration of the energy dispersion and the spin configuration 
	on the Fermi surface in (A)an FPFM and (B)ARM. 
	We assume that the exchange field is along the $z$-axis. 
	In addition, the RSOI $\lambda ({\bm \sigma} \times {\bm k})\cdot {\bm z}$ is considered in the ARM. 
	In the FPFM, the spin directions on the Fermi surface point to the $z$-direction, whereas, 
	in the ARM, they rotate along the Fermi surface and tilt to the $z$-axis. 
	}
	\label{fig:introduction}
\end{figure}
Determination of the pairing symmetry of the Cooper pair has been an important issue 
in the field of superconductivity. 
In this regard, tunneling spectroscopy is known to be useful. 
In the unconventional superconductor junctions, 
a zero-bias conductance peak (ZBCP) due to the surface Andreev bound state (SABS) is observed 
\cite{ABS,ABSb,Hu,kashiwaya00}, 
where the pair potential changes its sign on the Fermi surface
\cite{TK95,kashiwaya00}. 
Actually, the presence of a sharp ZBCP 
in the tunneling conductance in N/S junctions supports $d$-wave 
symmetry in cuprate
\cite{TK95}. 
In addition, a broad ZBCP observed in Sr$_{2}$RuO$_{4}$ junctions\cite{Kashiwaya11} 
is consistent with the SABS with linear dispersion such like 
chiral $p$-wave pairing
\cite{YTK97,YTK98,Honerkamp,IHSYMTS07}. 
Moreover, the tunneling spectroscopy in ferromagnet/superconductor (FM/S) junctions 
has also been studied up to now. 
For a spin-singlet superconductor, 
the magnitude of the tunneling conductance 
with the inner gap regime is suppressed\cite{Beenakker95}. 
In addition, in the case of a fully polarized ferromagnet metal (FPFM), 
the inner gap conductance is completely suppressed 
\cite{Kashiwaya99,Zutic1999,Zutic2000,Ting2000,Linder2010}. 
On the other hand, for a spin-triplet $p$-wave case \cite{Hirai2001,Hirai2003,Linder2010}, 
the resulting conductance depends on the direction of the $\bm{d}$-vector, 
which is perpendicular to the direction of the spin of spin-triplet Cooper pair. 

Recently, the role of the spin-orbit interactions on the tunneling spectroscopy 
in a superconductor has attracted much attention, 
potentially opening up a new direction for superconducting spintronics. 
Rashba spin-orbit interaction (RSOI) have a property to split the Fermi surface depending on the spin degrees of freedom, 
where the relative direction of the spin and momentum are locked owing to the RSOI in each Fermi surface
\cite{Rashba1960,Rashba1984,Molenkamp}. 
This unique property in a metal or doped semiconductor 
has attracted much attention in superconducting junctions as well as in the field of spintronics so far, 
since the direction of spin can be manipulated by the control of the RSOI
\cite{Streda,JinLian,Kato,Gundler,Sprisongmuang}. 
For example, 
the RSOI dependent charge transport has been studied 
in a two-dimensional electron gas (2DEG) with RSOI/$s$-wave superconductor junctions
\cite{Yokoyama2006,Yokoyama2009,Sun2015}. 

In the 2DEG, introducing an exchange field or applying an external magnetic field, 
a gap opens at the crossing point of two split bands by the RSOI\cite{Miron}. 
If we set the chemical potential in between the induced 
energy gap by manipulating the exchange field, 
the inner Fermi surface disappears. 
Thus, we can imagine novel quantum phenomena in the present system 
since only one of the Kramers doublet exists. 
In the following, we call this state an anomalous Rashba metal (ARM).
The aim of this paper is to study the tunneling spectroscopy in 
ARM/S junctions. 
A unique feature of the tunneling conductance is expected in ARM/S junctions 
owing to the reduction in spin degrees of freedom and the unique spin configuration of the ARM. 
Furthermore, it would be interesting to compare the ARM with the FPFM, 
both of which host a half of spin degrees of freedom; 
however, as shown in Fig. \ref{fig:introduction} (A) and (B), 
the spin textures in the band basis behave differently from each other. 
This difference gives a distinctive signature to each superconductor junction. 

Furthermore, it is known that the surface state of topological insulators (TIs)\cite{hasan10} 
also have a half of spin degrees of freedom and a unique spin texture\cite{Schwab}, 
which is the so-called helical metal. 
However, whereas TIs preserve time-reversal symmetry, ARMs break it. 
Thus, ARMs are fundamentally different from TIs. 
For superconductor junctions via a helical metal, there have been several studies 
on the surface of TIs and the unique feature of the charge transport in the systems has been reported
\cite{fu08,akhmerov09,law09,tanaka09,linder10}. 
While the properties of charge transport in the ARM/S junctions are naturally expected to be anomalous 
similar to the helical metal, 
they have not been revealed yet. 

In this paper, we theoretically study the tunneling conductance in the ARM/S junctions 
by solving the Bogoliubov-de Gennes (BdG) equation within the quasiclassical approximation 
for the several pairing symmetries: $s$-wave, $p$-wave, $d$-wave, chiral $p$-wave, 
helical $p$-wave, and chiral $d$-wave pairings. 
Among them, we reveal a qualitative difference between the N, the FPFM, and the ARM 
in superconducting junctions. 
For ARM/$s$-wave superconductor junctions, 
the magnitude of the inner gap conductance is enhanced 
as the RSOI increases; 
this behavior is clearly different from that of FPFM/$s$-wave superconductor junctions. 
In a similar manner, for ARM/$d_{xy}$-wave superconductor junctions, the RSOI retains 
the ZBCP. 
This contrasts sharply with the suppression of the ZBCP 
in FPFM/$d_{xy}$-wave superconductor junctions \cite{Kashiwaya99,Zutic1999,Zutic2000}. 
In addition, we find that, for ARM/$p_{x}$-wave superconductor junctions, 
the magnitude of the ZBCP significantly depends  
on the direction of the $\bm{d}$-vector in 
the $p_{x}$-wave superconductor. 
In our setup, the obtained ZBCP remains only 
when the $y$-component of the $\bm{d}$-vector is nonzero. 
This ${\bm d}$-vector dependance comes from the RSOI; 
thus, this feature is peculiar to ARM/$p_{x}$-wave superconductor junctions. 
We also show that the presence or absence of the ZCP is related to a topological number 
in $p_{x}$-wave superconductor. 
When the symmetry of S is 
a chiral $p$-wave, helical $p$-wave and chiral  
$d$-wave pairings, 
the pairing symmetries show qualitatively 
different line shapes of the tunneling conductance. 
 
The organization of this paper is as follows. 
In section \ref{formalism}, we explain our model and 
give a formulation of the tunneling 
conductance. 
In section \ref{swave}, the tunneling conductance of the ARM/$s$-wave superconductor 
junction is calculated. 
In section \ref{dwave}, the tunneling conductance of the ARM/$d_{xy}$-wave superconductor 
junction is shown.
We discuss the relevance to the tunneling spectroscopy of 
the LSMO/YBCO junction.
The calculation for the ARM/$p_{x}$-wave superconductor 
junction is shown in section \ref{pxandywave}. 
We interpret the obtained results 
using a chiral operator based on 
topology of the Hamiltonian. 
In section \ref{Sr2RuO4}, we show the tunneling conductance for 
ARM/chiral $p$-wave superconductor, ARM/chiral $d$-wave superconductor, 
and ARM/helical $p$-wave superconductor junctions. 
In section \ref{conclusion}, we conclude our results. 
\begin{figure}[b]
	\centering
	\includegraphics[width=5cm]{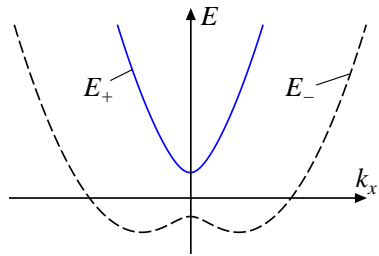}\\
	\caption{(Color online) The energy spectrum of the ARM.
					The eigenvalues are given by $E_{\pm} = {\xi}_{\bm{k}} \pm \sqrt{ {H}^{2} + (\lambda k)^{2} }$.}
	\label{fig:energydispersion}
\end{figure}
\section{Formulation for the tunneling conductance \label{formalism}}
Let us consider a two-dimensional ballistic ARM/insulator/superconductor 
junction in the ballistic limit. 
We assume that the ARM/S interface is located at $x = 0$ (along the $y$-axis). 
The interface has an infinitely narrow insulating barrier described by the delta function. 
In this section, a formulation of the tunneling conductance in the two-dimensional ARM/S junctions is shown. 

We start from the BdG Hamiltonian including both the exchange field and the RSOI as shown below, 
\begin{eqnarray}
	&& \ \ \ \ \bar{H} = \begin{bmatrix}                    \hat{H} ({\bm{k}})                      &  \hat{\Delta} ({\bm{k}}) \theta (x)  \\
	                                        {\hat{\Delta} ({\bm{k}})}^{\dagger} \theta (x) &    - {\hat{H} ({\bm{-k}})}^{\ast}  
	                 \end{bmatrix} , \label{eq:BdG} \\
	\hat{H} ({\bm{k}}) &=& \begin{bmatrix} {\xi}_{\bm{k}} + H \theta (-x) + V_{0} \delta (x) &            i \lambda k_{-} \theta (-x)                \\
						            - i \lambda k_{+} \theta (-x)                 & {\xi}_{\bm{k}} - H \theta (-x) + V_{0} \delta (x) 
				  \end{bmatrix} , \nonumber \\
				  \label{eq:NormalHamiltonian} \\
	&& \ \ \ \
	      \hat{\Delta} ({\bm{k}}) = i \hat{{\sigma}}_{y} ( {d}_{0} (\bm{k}) \hat{{\sigma}}_{0} + \bm{d} (\bm{k}) \hat{\bm{\sigma}} ) 
	                                     , \label{eq:PairPotential}
\end{eqnarray}
with $k_{\pm} = k_{x} \pm i k_{y}$, 
${\xi}_{\bm{k}} = \frac{k^{2}}{2m} - \mu_{N} \theta (-x) - \mu_{S} \theta (x)$, and $\hbar = 1$ . 
$\hat{\Delta} (\bm{k})$, 
$\mu_{N}$ ($\mu_{S}$), 
$\lambda (>0)$, $H (>0)$, and $\theta (x)$
are the pair potential, the chemical potential in the metal (superconductor), 
the amplitude of RSOI, 
the exchange field, 
and the step function, respectively. 
In Eq. (\ref{eq:PairPotential}), $d_{0} (\bm{k})$ denotes the pair potential in the spin-singlet superconductor, 
and $\bm{d} (\bm{k})$($=(d_{x}(\bm{k}), d_{y}(\bm{k}), d_{z}(\bm{k}))$) is the $\bm{d}$-vector of spin-triplet superconductor. 
When the spin-singlet (spin-triplet) superconductor is considered in $x > 0$, 
we choose $\bm{d} = {\bf{0}} ({d}_{0} = 0)$. 
Here, we assume that the exchange field is parallel to $z$-axis. 
Besides, the $z$-component of the RSOI $\lambda (\hat{{\bm \sigma}}\times\bm{k}) \cdot {\bm z}$ is considered, 
where ${\sigma}_{i} (i = 0, x, y, z)$ are the identity matrix and the Pauli matrix in the spin space. 
The energy spectrum of 
the ARM is given by $E_{\pm} = {\xi}_{\bm{k}} \pm \sqrt{ {H}^{2} + (\lambda k)^{2} }$ (see Fig.\ref{fig:energydispersion}). 
It should be made clear that our Hamiltonian is distinct from 
an ARM/spin-singlet $s$-wave superconductor hybrid system 
where the pair potential is induced in the ARM. 
In that case, the ARM hosts a chiral Majorana mode as  an edge state 
\cite{Alicea10,Jay,STF09,STF10,lutchyn10,oreg10,Yamakage2012}. 
In order to calculate the tunneling conductance of the ARM/S junctions, 
we choose $\left| \mu \right|< H$  in the following calculation. 
Fermi momenta for the outer(inner) Fermi surface $k_{1(2)}$ in the ARM is given as follows: 
\begin{eqnarray}
	&&k_{1 (2)} = \nonumber \\
	&&\sqrt{2m \biggl( {\mu}_{N} + m{\lambda}^{2} + (-) \sqrt{ {(m {\lambda}^{2})}^{2} + 2 m {\lambda}^{2} {\mu}_{N} +H^2 } \biggr) }. \label{eq:wavenumber} \nonumber  \\ 
\end{eqnarray}
Here, $k_{2}$ is a purely imaginary number and represents an evanescent wave because of the absence of the inner Fermi surface. 
To specify this, we define a real number ${\kappa}_{2}$ ($i {\kappa}_{2}=k_{2}$), 
\begin{eqnarray}
         {\kappa}_{2} = \sqrt{2m \biggl(  \sqrt{ {(m {\lambda}^{2})}^{2} + 2 m {\lambda}^{2} {\mu}_{N} +H^2 } - {\mu}_{N} - m{\lambda}^{2}  } \biggr). \nonumber  \\
\label{eq:evanescent}
\end{eqnarray}
From Eq. (\ref{eq:evanescent}), the $x$-component of $\kappa_{2}$ is given by 
\begin{eqnarray}
	{\kappa}_{2x} &=& \sqrt{{{\kappa}_{2}}^{2} + {k_{y}}^{2} }.
\end{eqnarray}
In the superconductor $(x>0)$,  the Fermi momentum $k_{S}$ can be denoted by 
$k_{S} \approx \sqrt{2m {\mu}_{S}}$ in the quasiclassical approximation. 
In addition, the $y$-component of all momenta satisfies 
\begin{eqnarray}
	k_{y} = k_{1} \sin {\theta}_{N} = k_{S} \sin {\theta}_{S} ,
\end{eqnarray}
because a momentum 
parallel to the interface is conserved 
when we assume a flat interface. 

\begin{figure}[t]
	\centering 
	\includegraphics[width=6cm]{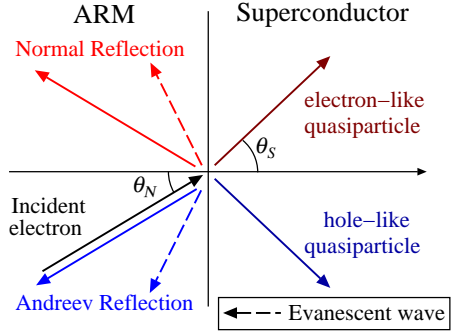}\\
	\caption{(Color online) Schematic illustration of the scattering process. ${\theta}_{N}$ is an incident angle of $k_{1}$ 
	respect to the interface normal. 
	${\theta}_{S}$ denotes the direction of motions of quasiparticles in S measured from the interface normal.}
	\label{fig:Scattering}
\end{figure}
\begin{figure}[t]
	\centering 
	\includegraphics[width=5cm]{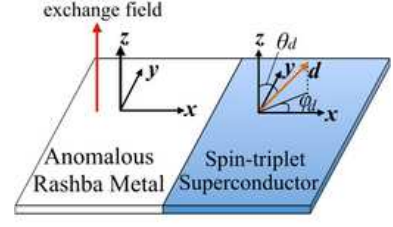}
	\caption{(Color online) Schematic illustration of 
			an ARM/$p$-wave superconductor junction. 
			The exchange field is parallel to $z$-axis.}
	\label{fig:vector}
\end{figure}
First, we introduce a wave function in the ARM. 
As shown in Fig. \ref{fig:Scattering}, the wave function $\psi (x,y)$ in the ARM is represented by using eigenfunctions of the Hamiltonian. 
\begin{eqnarray}
	\psi (x>0,y) &=&e^{i k_{y}y } \Biggl(  
	                  e^{i k_{1} \cos {\theta}_{N} x}           \begin{bmatrix}            s        \\ 1 \\           0           \\  0  \end{bmatrix} \nonumber \\
	     &+& r_{1} e^{-i k_{1} \cos {\theta}_{N} x}          \begin{bmatrix}      {s}^{\ast} \\ 1 \\           0           \\  0  \end{bmatrix}
	       +   a_{1} e^{i k_{1} \cos {\theta}_{N} x}           \begin{bmatrix}             0        \\ 0 \\   - {s}^{\ast}     \\  1 \end{bmatrix} \nonumber \\
	     &+& r_{2} e^{{\kappa}_{2 x} x}  \begin{bmatrix}           t_{e}    \\ 1 \\          0            \\  0  \end{bmatrix}
	       +   a_{2} e^{{\kappa}_{2 x} x}  \begin{bmatrix}             0        \\ 0 \\       t_{h}         \\  1 \end{bmatrix}
 					    \Biggr) , \\
	&&    s         = -\frac{i \lambda k_{1} e^{-i{\theta}_{N} }}{{\xi}_{{\bm k}_{1}}  + H} , \\
	&&{t}_{e}     = -\frac{\lambda (  \kappa_{2 x} + k_{y})}{\xi_{i {\bm \kappa}_{2}} + H} , \\ 
	&&{t}_{h}     =  \frac{\lambda (- \kappa_{2 x} + k_{y})}{\xi_{i {\bm \kappa}_{2}} + H} ,
\end{eqnarray}
where $r_{1}$ and $r_{2}$ ($a_{1}$ and $a_{2}$) are normal (Andreev) reflection coefficients and 
${\theta}_{N}$ is an injection angle of $k_{1}$ measured from the 
normal to the interface (see Fig. \ref{fig:Scattering}). 
In addition, we assume ${\mu}_{S} \pm {\Delta}_{0} \approx {\mu}_{S}$ in the quasiclassical approximation. 
An injected electron can not transmit into the superconductor 
for ${\theta}_{N} > \arcsin(\frac{k_{S}}{k_{1}}) (\equiv {\theta}_{C})$. 
Next, we calculate a wave function in the superconductors. 
With the magnitude of the pair potential ${\Delta}_{0}$, the pair potential matrices for 
spin-singlet and spin-triplet superconductors are given by 
\begin{eqnarray}
\hat{\Delta} (\bm{k}) = \begin{cases}
	\begin{bmatrix}
		                        0       &  {d}_{0} ({\bm{k}})   \\
		 - {d}_{0} ({\bm{k}})   &             0                               \end{bmatrix} , \ \ \ \ \ \ \ \ (\text{spin-singlet pair}) & \\ \\
		 
	\begin{bmatrix}
		 - d_{x} (\bm{k}) +  i d_{y}(\bm{k}) &               d_{z}(\bm{k})           \\
		                 d_{z}(\bm{k})                & d_{x}(\bm{k}) + i d_{y}(\bm{k})  \end{bmatrix}.  
		                 \\ \ \ \ \ \ \ \ \ \ \ \ \ \ \ \ \ \ \ \ \ \ \ \ \ \ \ \ \ \ \ \ \ \ \  (\text{spin-triplet pair}) & 
		                 \end{cases} \label{eq:PP}
\end{eqnarray}
In Eq.(\ref{eq:PP}), $d_{0} (\bm{k})$ is defined as $d_{0} (\bm{k}) \equiv {\Delta}_{0} f_{\theta_{S}}$, 
where $f_{\theta_{S}}$ denotes the momentum dependance of the pair potential on the Fermi surface in spin-singlet superconductor. 
The direction of the ${\bm d}$-vector is denoted by the polar angle ${\theta}_{d}$ and the azimuthal angle ${\phi}_{d}$ in Fig. \ref{fig:vector}. 
The $\bm{d}$-vector for the $p_{x}$-wave, $p_{y}$-wave, or 
chiral $p$-wave superconductor is given by 
\begin{eqnarray}
	\bm{d} &=& (d_{x} , d_{y} , d_{z} ) \nonumber \\
		   &=& {\Delta}_{0} g_{{\theta}_{S}} ( \sin {\theta}_{d} \cos {\phi}_{d} ,
		          					       \sin {\theta}_{d} \sin {\phi}_{d} ,
								       \cos {\theta}_{d} ). \nonumber \\ \label{eq:dvectoreqnonhelical}
\end{eqnarray}
In addition, we assume that the $\bm{d}$-vector for the helical $p$-wave superconductor is given by 
\begin{eqnarray}
	\bm{d} = {\Delta}_{0} ( w_{1 {\theta}_{S}},
					  w_{2 {\theta}_{S}},
					  0 ). \label{eq:dvectoreqhelical}
\end{eqnarray}
Similar to $f_{\theta_{S}} $, $g_{\theta_{S}}$ and $w_{i \theta_{S}}$ ($i=1, 2$) represent 
the momentum dependance of the pair potential on the Fermi surface in the spin-triplet superconductors. 
The explicit form of  $f_{\theta_{S}}$, $g_{\theta_{S}}$, $w_{i {\theta}_{S}} (i = 1, 2)$ are 
given in sections \ref{Result} and \ref{Sr2RuO4}. 
The wave functions in the spin-singlet and 
spin-triplet superconductors are given as follows:\\
(i) spin-singlet superconductor 
\begin{eqnarray}
	\psi (x,y) &=& e^{i k_{y}y} \Biggl(
	          s_{1} e^{ i k_{FS} \cos {\theta}_{S} x}  \begin{bmatrix}                       1              \\                   0                     \\              0           \\ {\Gamma}_{+} \end{bmatrix}  \nonumber \\ 
	  &+& s_{2} e^{ i k_{FS} \cos {\theta}_{S} x}  \begin{bmatrix}                       0               \\                  1                     \\ - {\Gamma}_{+} \\         0              \end{bmatrix} \nonumber \\
	  &+& s_{3} e^{-i k_{FS} \cos {\theta}_{S} x}  \begin{bmatrix}                       0               \\   - {{\tilde{\Gamma}}}_{-}  \\               1          \\         0              \end{bmatrix} \nonumber \\
	  &+& s_{4} e^{-i k_{FS} \cos {\theta}_{S} x}  \begin{bmatrix}     {\tilde{\Gamma}}_{-}    \\                 0                      \\              0          \\         1              \end{bmatrix}
 					    \Biggr) , \\
		 {\Gamma}_{+} &=& \frac{\Delta_{0} {f_{{\theta}_{S}}}^{\ast} }{E + \sqrt{E^{2} - {{\Delta}_{0}}^{2}    |f_{{\theta}_{S}}|^{2}     }} ,    \\
	\tilde{{\Gamma}}_{-}  &=& \frac{\Delta_{0}     f_{\pi - {\theta}_{S}}  }{E + \sqrt{E^{2} - {{\Delta}_{0}}^{2} |f_{\pi - {\theta}_{S}}|^{2} }},
\end{eqnarray} 
(ii) $p_{x}$-wave, $p_{y}$-wave, and chiral $p$-wave superconductors
\begin{eqnarray}
	\psi (x,y) &=& 
	                  e^{i k_{y} y}  
	           \Biggl(  
	          s_{1} e^{ i k_{FS} \cos {\theta}_{S} x}  \begin{bmatrix}            - B             \\    \cos {\theta}_{d}     \\   {\Gamma}_{+}   \\    0           \end{bmatrix} \nonumber \\
	  &+& s_{2} e^{ i k_{FS} \cos {\theta}_{S} x}  \begin{bmatrix}  \cos {\theta}_{d}  \\          B^{\ast}            \\     0     \\    {\Gamma}_{+}       \end{bmatrix} \nonumber \\
          &+& s_{3} e^{-i k_{FS} \cos {\theta}_{S} x}  \begin{bmatrix}  \tilde{{\Gamma}}_{-}   \\    0    \\     - B^{\ast}       \\  \cos {\theta}_{d}    \end{bmatrix} \nonumber \\
	  &+& s_{4} e^{-i k_{FS} \cos {\theta}_{S} x}  \begin{bmatrix}     0    \\  \tilde{{\Gamma}}_{-}  \\  \cos {\theta}_{d} \\           B                \end{bmatrix} 
 					    \Biggr) ,  \\
	&B& = \sin {\theta}_{d} \cos {\phi}_{d} + i \sin {\theta}_{d} \sin {\phi}_{d} , \\
		 {\Gamma}_{+} &=& \frac{ \Delta_{0} {g_{{\theta}_{S}}}^{\ast}}{E + \sqrt{E^{2} - {{\Delta}_{0}}^{2}  |g_{{\theta}_{S}}|^{2}}} ,    \\
	\tilde{{\Gamma}}_{-} &=& \frac{ \Delta_{0}  g_{\pi - {\theta}_{S}}    }{E + \sqrt{E^{2} - {{\Delta}_{0}}^{2}   |g_{\pi - {\theta}_{S}}|^{2}}},
\end{eqnarray}
(iii) helical $p$-wave superconductor
\begin{eqnarray}
	\psi (x,y) &=& 
	                  e^{i k_{y} y}
	           \Biggl(
	               s_{1} e^{ i k_{FS} \cos {\theta}_{S} x}  \begin{bmatrix}      1    \\   0   \\       - {\Gamma}_{1+} + i {\Gamma}_{2+} \\ 0  \end{bmatrix} \nonumber \\
	       &+& s_{2} e^{ i k_{FS} \cos {\theta}_{S} x}  \begin{bmatrix}      0    \\  - 1  \\  0  \\ {\Gamma}_{1+} + i {\Gamma}_{2+} \end{bmatrix} \nonumber \\
               &+& s_{3} e^{-i k_{FS} \cos {\theta}_{S} x}  \begin{bmatrix}   - ( \tilde{{\Gamma}}_{1-} + i \tilde{{\Gamma}}_{2-} ) \\ 0 \\ 1 \\  0  \end{bmatrix} \nonumber \\
	       &+& s_{4} e^{-i k_{FS} \cos {\theta}_{S} x}  \begin{bmatrix}      0 \\ \tilde{{\Gamma}}_{1-} + i \tilde{{\Gamma}}_{2-} \\  0  \\  1   \end{bmatrix}
 					    \Biggr) ,  \\ 
		 {\Gamma}_{j +} &=& \frac{ \Delta_{0} {w_{j {\theta}_{S}}}^{\ast}}{E + \sqrt{E^{2} - {{\Delta}_{0}}^{2}  |w_{j {\theta}_{S}}|^{2}}} ,    \\ 
	\tilde{{\Gamma}}_{j -} &=& \frac{ \Delta_{0}  w_{j \pi - {\theta}_{S}}    }{E + \sqrt{E^{2} - {{\Delta}_{0}}^{2}  |w_{j \pi - {\theta}_{S}}|^{2}}}. 
\end{eqnarray}
In the above, $s_{l}$ ($l=1, 2, 3 ,4$) is the transmission coefficients  
and $j=1, 2$.
${\theta}_{S}$ is the angle of the momentum $k_{S}$ with respect to the interface normal (see Fig. \ref{fig:Scattering}). 
Since we assume that the wave function in the junction is continuous at the interface, 
the boundary conditions is given as follows: 
\begin{eqnarray}
	&&\ \ \ \ \ \ \ \ \ \ \ \ \ \psi (+0,y) - \psi (-0,y) = 0  , \\
	&&\bar{{v}_{x}} (\psi (+0 ,y) - \psi (-0,y)) = \frac{1}{m i} 2 m V_{0} \hat{{\sigma}_{0}} \hat{{\tau}_{z}} \psi (0,y) , \nonumber \\\label{eq:boundary}
\end{eqnarray}
where ${\hat{\sigma}}_i$ $({\hat{\tau}}_i)$ ($i=0,x,y,z$) are the identity matrix and the Pauli matrices in the spin (Nambu) space. 
In Eq. (\ref{eq:boundary}), the velocity operator in the $x$-direction ${\bar{v}_{x}}$ is defined by\cite{Molenkamp} 
\begin{eqnarray}
	{\bar{v}}_{x} = \frac{\partial \bar{H}}{\partial k_{x}} = \begin{bmatrix}   \frac{1}{m i} \frac{\partial}{\partial x}   &            i \lambda \theta (-x)               &   0    &    0     \\
						                                                                             - i \lambda \theta (-x)               & \frac{1}{m i} \frac{\partial}{\partial x} &   0    &    0     \\
						                                                             0    &   0   & - \frac{1}{m i} \frac{\partial}{\partial x}  &           - i \lambda \theta (-x)                \\
						                                                             0    &   0   &             i \lambda \theta (-x)                 & - \frac{1}{m i} \frac{\partial}{\partial x}      
			                                                        \end{bmatrix} . \nonumber \\
\end{eqnarray}
By solving Eq. (\ref{eq:boundary}), we determine $a_{1}$ and $b_{1}$ and 
obtain the normalized tunneling conductance\cite{Molenkamp}, 
\begin{eqnarray}
	\sigma (eV) &=& \frac{\int^{{\theta}_{C}}_{-{\theta}_{C}} {\sigma}_{S} (eV,{\theta}_{S}) d {\theta}_{S}}
					{\int^{{\theta}_{C}}_{-{\theta}_{C}} {\sigma}_{N} (eV,{\theta}_{S}) d {\theta}_{S}}, \label{eq:normcond} \\
	{\sigma}_{S} (eV,{\theta}_{S}) &=& 4e (1 + {|a_{1}|}^{2} - {|r_{1}|}^{2}) \nonumber \\ 
						&\times& \bigl( \frac{{k}_{1} \cos {\theta}_{N}}{m} ({|s|}^{2} + 1 ) - i \lambda ( s - {s}^{\ast}) \bigr)  
						\label{eq:superconductor}
\end{eqnarray}
${\sigma}_{S}$ (${\sigma}_{N}$) represents the tunneling conductance in the ARM/S junction (the ARM/normal metal (${\Delta}_{0}=0$) junction). 
In section \ref{Result}, we also show the tunneling conductance of one-dimensional limit by choosing $k_{y} = 0$. 
\section{Results and Discussions\label{Result}}
In this and the next sections, we show and discuss the obtained tunneling conductance of ARM/S junctions for various types of 
a pairing symmetry, 
where dimensionless parameters, $\alpha = \frac{m {\lambda}^{2} }{ {\mu}_{S} }$, 
$\gamma = \frac{{\mu}_{N}}{{\mu}_{S}}$, $h = \frac{ H }{ {\mu}_{S} }$, 
and $Z = \frac{V_{0} k_{S}}{ {\mu}_{S}}$ are used. 
For simplicity, we use abbreviations for superconducting junctions, 
$e.g.$, ARM/$s$-waves and ARM/spin-singlets in the following sections. 

\subsection{ARM/$s$-wave superconductor junction \label{swave}}
\begin{figure}[t]
		\centering
		\includegraphics[width=8cm]{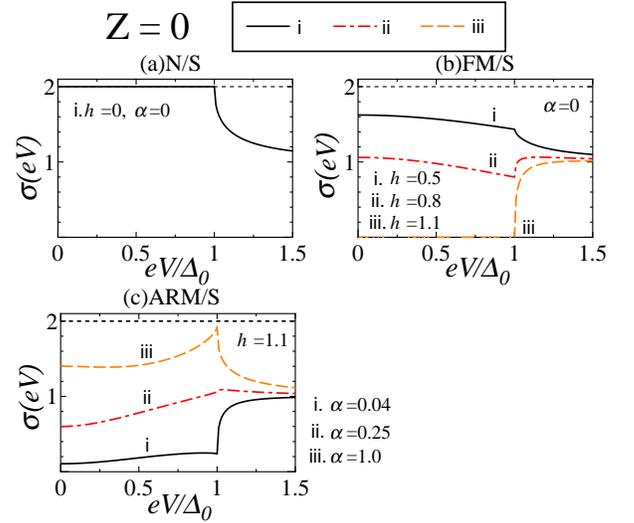}\\
		\caption{(Color online) Normalized tunneling conductance $\sigma(eV)$ of two-dimensional 
		(a)N/S, (b)FM/S, and (c)ARM/S junctions without insulating barrier ($Z=0$), 
		where S is chosen as the $s$-wave superconductor. 
		We use $\gamma = 1.0$ in all cases. 
		 }
		\label{fig:2Dswave}
	\end{figure}

In this subsection, we discuss 
two-dimensional ARM/s-wave superconductor junctions with  
\begin{eqnarray}
	f_{{\theta}_{S}} = 1. 
\end{eqnarray}
We calculate the normalized tunneling conductance $\sigma(eV)$ 
in Eq. (\ref{eq:normcond}) using the formulation 
in section \ref{formalism}. 
First, we show the obtained conductance without an insulating barrier, $i.e.$, $Z=0$, as a function of bias voltage. 
Figure \ref{fig:2Dswave} shows the $\sigma(eV)$ of 
an N/$s$-wave (a), 
FM/$s$-waves (b), and ARM/$s$-waves (c) for $\gamma=1$. 
In Fig. \ref{fig:2Dswave} (a), we find $\sigma(|eV|<{\Delta}_{0})=2$ 
by the perfect Andreev reflection at the interface \cite{Andreev,BTK}. 
As shown in Fig. \ref{fig:2Dswave}(b), in FM/S junctions, 
the $\sigma(eV)$ are shown for 
various magnitude of the exchange field, namely $h$. 
The magnitude of the inner gap conductance $\sigma(|eV|<\Delta_{0})$ is 
suppressed with the increase in $h$. 
Especially, the $\sigma(|eV|<{\Delta}_{0})$ becomes zero for $h>1$, 
where  the ferromagnet is fully polarized (see Fig. \ref{fig:2Dswave}(b)(iii)). 
As shown in  Fig. \ref{fig:2Dswave}(c), the $\sigma(eV)$ for $h>1$ 
is enhanced with the increase in the 
magnitude of the RSOI  $\alpha$.
Qualitative features of the $\sigma(eV)$ in Fig. \ref{fig:2Dswave}(c) 
can be interpreted by the spin configuration of the ARM. 

\begin{figure}[t]
		\centering
		\includegraphics[width=8cm]{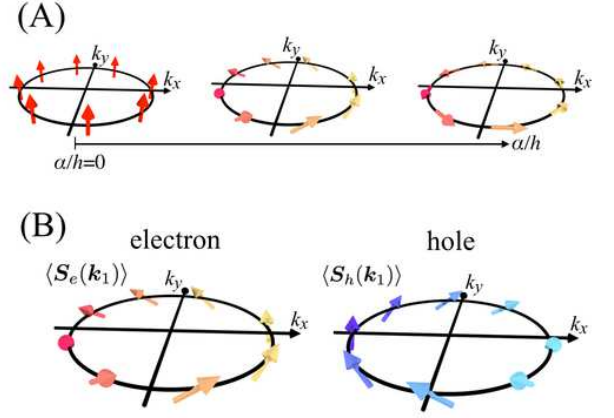}\\
		\caption{(Color online) Schematic illustration of the spin configuration in the ARM. 
		(A) $\alpha/h$ dependance of the spin configuration and 
		(B) the spin configuration of an electron and hole is shown.}
		\label{fig:spintexture}
	\end{figure}

\begin{figure}[t]
		\centering
		\includegraphics[width=5cm]{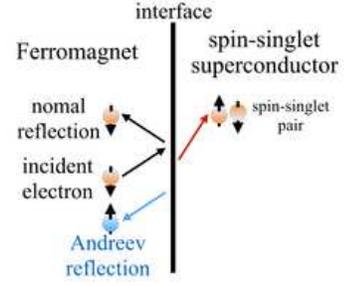}\\
		\caption{(Color online) Schematic illustration of 
		the scattering process at the interface of FPFM/spin-singlet 
		superconductor junctions. }
		\label{fig:spinsinglet}
	\end{figure}

To see this, we calculate the spin configuration in the ARM. 
Using the eigenfunction of the ARM 
$\psi ({\bm k}_{1}) = 
{(\frac{s_{{\bm k}_{1}}}{\sqrt{{|s_{{\bm k}_{1}}|}^{2} + 1}},\frac{1}{\sqrt{{|s_{{\bm k}_{1}}|}^{2} + 1}})}^{T}$, 
the spin direction of electron and hole states is defined by 
$\langle {\bm{S}}_{e}({\bm k}_{1})\rangle \equiv 
\langle \psi ({\bm k}_{1}) | \hat{{\bm \sigma}} | \psi({\bm k}_{1}) \rangle$ 
and 
$\langle {\bm{S}}_{h}({\bm k}_{1})\rangle \equiv 
   \langle {\psi(-{\bm k}_{1})}^{\ast} | {\hat{{\bm \sigma}}}^{\ast} | {\psi(-{\bm k}_{1})}^{\ast} \rangle$, 
respectively. 
In the above, 
$s_{{\bm k}_{1}}$ is given by $s_{{\bm k}_{1}} = - \frac{\lambda (ik_{1x} + k_{1y})}{\xi_{{\bm k}_{1}} + H}$, 
where $k_{1x(y)}$ is a $x$ ($y$)-component of ${{\bm k}}_{1}$. 
The explicit forms of 
$\langle {{\bm S}}_{e}({\bm k}_{1})\rangle$ and $\langle {{\bm S}}_{h}({\bm k}_{1})\rangle$
become 
\begin{eqnarray}
	&&\langle{\bm{S}}_{e}({\bm k}_{1})\rangle = 
	( \langle{S}_{xe}({\bm k}_{1})\rangle, 
	  \langle{S}_{ye}({\bm k}_{1})\rangle, 
	  \langle{S}_{ze}({\bm k}_{1})\rangle ) \nonumber \\
	&=& 
	\Bigl(\frac{ -2 \lambda k_{1y} {\epsilon}_{\bm{k_{1}}} }{ (\lambda k_{1})^{2} + {{\epsilon}_{{\bm k}_{1}}}^{2}}   \ , \
		\frac{ 2 \lambda k_{1x} {\epsilon}_{\bm{k_{1}}}}{ (\lambda k_{1})^{2} + {{\epsilon}_{{\bm k}_{1}}}^{2}}  \ , \
		\frac{(\lambda k_{1})^{2} - {{\epsilon}_{\bm{k_{1}}}}^{2} }{ (\lambda k_{1})^{2} + {{\epsilon}_{{\bm k}_{1}}}^{2}} \Bigr), \nonumber \\ \label{eq:electronspin}  
\end{eqnarray}
\begin{eqnarray}
	&&\langle{\bm{S}}_{h}({\bm k}_{1})\rangle = 
	( \langle{S}_{xh}({\bm k}_{1})\rangle, 
	  \langle{S}_{yh}({\bm k}_{1})\rangle, 
	  \langle{S}_{zh}({\bm k}_{1})\rangle ) \nonumber \\
	&=& 
	\Bigl(\frac{ 2 \lambda k_{1y} {\epsilon}_{- \bm{k_{1}}} }{ (\lambda k_{1})^{2} + {{\epsilon}_{-{\bm k}_{1}}}^{2}}   \ , \
		\frac{-2 \lambda k_{1x} {\epsilon}_{- \bm{k_{1}}}}{ (\lambda k_{1})^{2} + {{\epsilon}_{-{\bm k}_{1}}}^{2}}  \ , \
		\frac{(\lambda k_{1})^{2} - {{\epsilon}_{- \bm{k_{1}}}}^{2} }{ (\lambda k_{1})^{2} + {{\epsilon}_{-{\bm k}_{1}}}^{2}} \Bigr), \nonumber \\ \label{eq:holespin}
\end{eqnarray}
with
\begin{eqnarray}
			{\epsilon}_{{\bm k}_{1}} = {\xi}_{\bm{k_{1}}} + H .
\end{eqnarray}
If we choose $\lambda=0$, we can reproduce the FPFM case. 
From Eqs. (\ref{eq:wavenumber}), (\ref{eq:electronspin}), and (\ref{eq:holespin}), 
while the sign of the in-plane components of each spin expectation value are opposite, 
the $z$-component of those is the same: 
\begin{eqnarray}
	\langle{S}_{ze}({\bm k}_{1})\rangle &=& \langle{S}_{zh}({\bm k}_{1})\rangle  \nonumber \\
					&=& -\frac{1}{1+ \frac{2\alpha}{h} + \frac{2\alpha(1 - h)}{h (h + \alpha + \sqrt{{\alpha}^{2} + 2\alpha + h^{2}}) } }. \label{spinapproachzro}
\end{eqnarray}
For Eq. (\ref{spinapproachzro}), we can find that, if $\alpha \gg h$, 
$\langle{S}_{ze}({\bm k}_{1})\rangle$ and $\langle{S}_{zh}({\bm k}_{1})\rangle$ 
approach zero (see Fig. \ref{fig:spintexture}(A)). 
On the other hand, the magnitudes of 
$\langle{S}_{x(y)e}({\bm k}_{1})\rangle$ and $\langle{S}_{x(y)h}({\bm k}_{1})\rangle$
become larger as the magnitude of the RSOI increases. 
These indicate that the spin in the ARM is not fully polarized along $z$-axis and 
its direction has an $xy$-plane component unlike the FPFM. 
We show later that 
$x$ and $y$-components of the spin polarization 
induced by the RSOI do not suppress the 
magnitude of the $\sigma(|eV|<{\Delta}_{0})$ in the ARM/spin-singlet. 
As a preparation for showing it, we explain why the tunneling conductance in FPFM/spin-singlets 
is reduced by the exchange field. 
Figure \ref{fig:spinsinglet} shows the scattering process where 
an electron with down-spin is injected from the left side. 
In this case, the spin of an incident electron is flipped 
through the Andreev reflection because 
we assume the spin-singlet superconductor for $x>0$. 
However, the Andreev reflection for $|eV|<{\Delta}_{0}$ 
does not occur in the FPFM/spin-singlets 
since there is no corresponding Fermi surface for the hole state with up-spin. 
Equations (\ref{eq:electronspin}) and (\ref{eq:holespin}) confirm this since 
$\langle{S}_{ze}({\bm k}_{1})\rangle = \langle{S}_{zh}({\bm k}_{1})\rangle=-1$ 
is satisfied in the FPFM. 
This is because the Andreev reflection is suppressed in FPFM/spin-singlets. 
In addition, the suppression of the Andreev reflection reduces the inner gap conductance 
as we can see from 
\begin{eqnarray}
	\sigma(eV) \propto 1 - {|r|}^{2} + {|a|}^{2}, 
\end{eqnarray}
where $r$($a$) is a normal(Andreev) reflection coefficient. 
Therefore, the tunneling conductance decreases 
because of the exchange field in FPFM/spin-singlets. 
On the other hand, for $\lambda \neq 0$, $\langle{S}_{x(y)e}({\bm k}_{1})\rangle$ and 
$\langle{S}_{x(y)h}({\bm k}_{1})\rangle$ become nonzero and satisfy 
\begin{eqnarray}
	\langle S_{xe}({\bm k}_{1})\rangle &=& - \langle S_{xh}({\bm k}_{1})\rangle ,\\
	\langle S_{ye}({\bm k}_{1})\rangle &=& - \langle S_{yh}({\bm k}_{1})\rangle.
\end{eqnarray}
in the ARM as shown in Fig. \ref{fig:spintexture}(B). 
This means that the coefficient of the Andreev reflection 
recovers owing to the RSOI in the ARM/spin-singlets by the comparison with the 
FPFM/spin-singlets. 
Accordingly, the presence of the RSOI enhances 
the magnitude of the inner gap conductance in ARM/spin-singlets. 
The above  explanation is consistent with 
the results in Fig. \ref{fig:2Dswave}(c). 
	\begin{figure}[t]
		\centering
		\includegraphics[width=5cm]{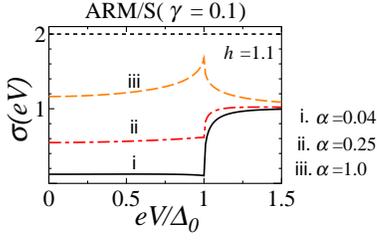}\\
		\caption{(Color online) Normalized tunneling conductance $\sigma(eV)$ of two-dimensional 
		ARM/S junctions  without insulating barrier ($Z=0$), 
		where S is chosen as the $s$-wave superconductor. 
		We use $\gamma = 0.1$ in all cases. }
		\label{fig:2Dswave2}
	\end{figure}

Also, we calculate the $\sigma (eV)$ of ARM/$s$-waves with $\gamma=0.1$ 
because $\gamma<1$ should be satisfied in realistic cases. 
The results are shown in Fig.\ref{fig:2Dswave2}. 
Since the $\sigma(|eV|<{\Delta}_{0})$ in Fig.\ref{fig:2Dswave2} is enhanced with the increase in $\alpha$, 
it is found that the change of $\gamma$ does not qualitatively influence the feature of the $\sigma(eV)$ in ARM/$s$-waves. 
In addition, even for junctions with the anisotropic superconductor, 
the qualitative features of $\sigma(eV)$ are insensitive to the change of $\gamma$. 
So, we mainly study for $\gamma=1$ below. 

\begin{figure}[t]
		\centering
		\includegraphics[width=8cm]{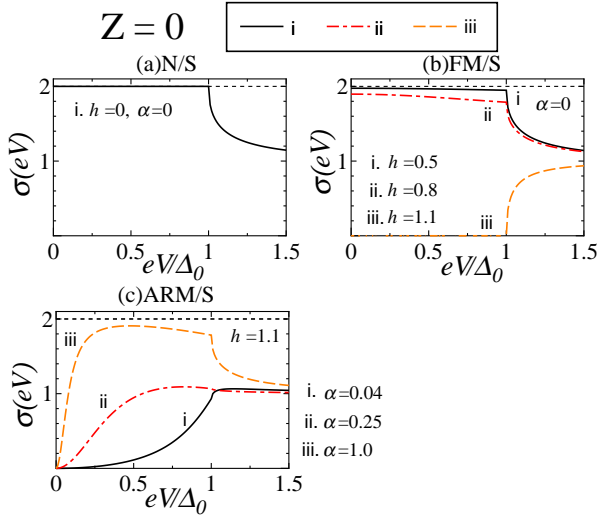}\\
		\caption{(Color online) Normalized $\sigma(eV)$ of 
		(a)N/S, (b)FM/S, and (c)ARM/S junctions without insulating barrier ($Z=0$) 
		in one-dimensional limit, 
		where S is chosen as the $s$-wave superconductor. 
		We use $\gamma = 1.0$ in all cases. 
		 }
		\label{fig:1Dswave}
	\end{figure}

Next, tunneling conductance in the one-dimensional limit 
which corresponds to the angle resolved conductance with 
perpendicular injection ($k_{y}=0$) is studied.  
Figure \ref{fig:1Dswave} shows 
$\sigma(eV)$ of the one-dimensional system
for $Z=0$. 
The indices of Figs. \ref{fig:1Dswave} (a), (b), and (c) correspond with those of Fig. \ref{fig:2Dswave}. 
As we can see from Figs. \ref{fig:1Dswave}(a) and \ref{fig:1Dswave}(b), for an N/$s$-wave and FM/$s$-waves, 
the qualitative behaviors of the $\sigma(eV)$ in one-dimensional limit are similar to those in two-dimensional cases
\cite{Andreev,BTK}. 
Figure \ref{fig:1Dswave}(c) also indicates that the $\sigma(|eV|<{\Delta}_{0})$ 
increases owing to the RSOI 
(see Fig. \ref{fig:1Dswave}(b)iii). 
However, note that zero bias conductance (ZBC), $i.e.$, 
$\sigma(eV=0)$, is 
zero regardless of the change of $\alpha$. 
This is because, in the one-dimensional cases, $|{a}_{1}|^{2}=0$ and $|{r}_{1}|^{2}=1$ are satisfied for $eV=0$ in the Eq.\ref{eq:superconductor}. 
This profile of the $\sigma(eV=0)$ does not correspond with that 
in the corresponding two-dimensional cases, 
but the result is consistent with the previous works\cite{Beri2009,Ioselevich2013,San-Jose}. 
According to one of the previous works\cite{Ioselevich2013}, 
where the conductance is calculated by the scattering matrix theory, 
the ZBC should be quantized to be $0$ or $2$ 
if the half of spin degrees of freedom 
and one-channel system are realized in the normal metallic region. 
Moreover, if the superconductor in the junction is topologically trivial, 
ZBC should be zero\cite{Ioselevich2013}. 
In our model, the one-dimensional ARM is just a one-channel system, and 
we consider the topologically trivial $s$-wave superconductor in $x > 0$. 
Hence, the ZBC should be zero in the one-dimensional ARM/$s$-waves. 

\begin{figure}[t]
	 	\centering
		\includegraphics[width=8cm]{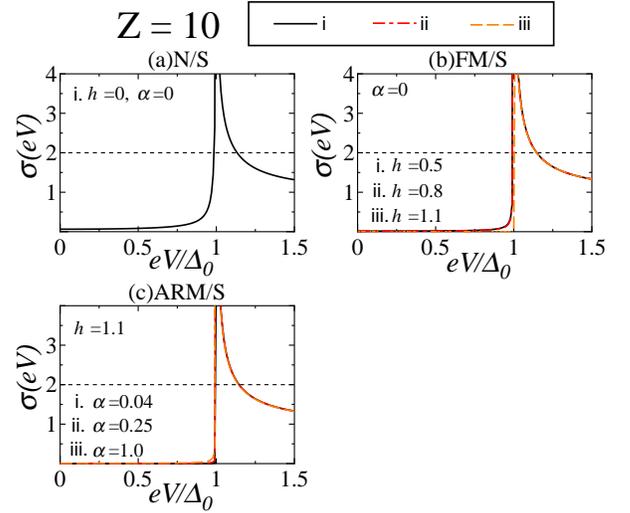}\\
		\caption{(Color online) Normalized $\sigma(eV)$ of two-dimensional
		(a)N/S, (b)FM/S, and (c)ARM/S junctions with high tunneling barrier($Z=10$), 
		where S is chosen as $s$-wave superconductor. 
		We use $\gamma = 1.0$ in all cases. 
		 }
		\label{fig:2Dswavebarrior}
	\end{figure}

Then, we show tunneling conductance with 
high-barrier case ($Z=10$) for the two-dimensional junctions. 
$\sigma(eV)$ of an N/$s$-wave, 
FM/$s$-waves and ARM/$s$-waves are 
plotted in Figs. \ref{fig:2Dswavebarrior}(a), 
\ref{fig:2Dswavebarrior}(b) and \ref{fig:2Dswavebarrior}(c), 
respectively. 
In these cases, all of the line shapes of the $\sigma(eV)$ show conventional U-shaped structures 
regardless of the change of $\alpha$ and $h$ (see Fig. \ref{fig:2Dswavebarrior}) 
since the $\sigma(|eV|<{\Delta}_{0})$ is strongly reduced 
by the insulating barrier due to the absence of the SABS. 
Namely, the coexistence of the exchange field and the RSOI does not 
qualitatively affect the $\sigma(eV)$ for the high-barrier case. 
\subsection{ARM/$d$-wave superconductor junction \label{dwave}}
In order to understand the effect of an SABS 
\cite{TK95,Hu} on the charge transport of ARM/S junctions, 
we calculate the tunneling conductance in 
two-dimensional ARM/$d$-wave superconductor junctions in this subsection. 
As a typical example of $d$-wave superconductor, 
we choose the $d_{{x}^{2} -  {y}^{2}}$-wave and $d_{xy}$-wave pair potentials.
In these cases, $f_{{\theta}_{S}}$ is given by 
	\begin{eqnarray}
		f_{{\theta}_{S}} =
		\begin{cases}
			\cos(2 {\theta}_{S} )  & (d_{{x}^{2} -  {y}^{2}}\text{-wave}) \\
			\sin(2 {\theta}_{S} )   & (d_{xy}\text{-wave})
		\end{cases}.
	\end{eqnarray}

First, using $f_{\theta_{S}}$, tunneling conductance for $Z=0$ is studied. 
It is known that, 
in FM/$d$-waves, the inner gap conductance is suppressed by the exchange field, 
and the ZBC becomes zero 
when the ferromagnet is fully polarized. 
As we have discussed in section \ref{swave}, in ARM/spin-singlets, 
the inner gap conductance recovers with increasing the magnitude of the RSOI. 
ARM/$d$-waves also show the enhancement of the inner gap conductance due to the RSOI. 
In addition, the qualitative features of the tunneling conductance 
does not depend on whether the paring symmetry is 
$d_{x^{2}-y^{2}}$-wave or $d_{xy}$-wave. 

Next, we focus on the tunneling conductance for the high-barrier case ($Z=10$). 
The line shape of the $\sigma(eV)$ becomes the conventional $V$-shaped 
structure for $d_{{x}^{2} -  {y}^{2}}$-wave 
superconductor junctions regardless of the change of $\alpha$ and $h$. 
This is because the $\sigma(eV)$ is strongly reduced 
by the insulating barrier. 

\begin{figure}[t]
		\centering
		\includegraphics[width=8cm]{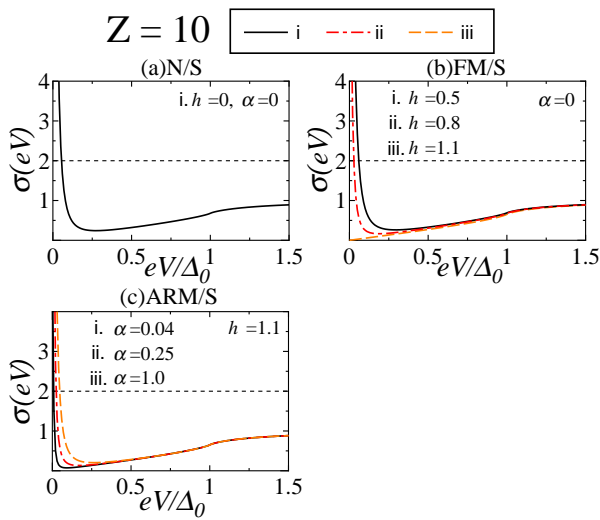}\\
		\caption{(Color online) Normalized $\sigma(eV)$ of two-dimensional
		(a)N/S, (b)FM/S, and (c)ARM/S junctions with high tunneling barrier ($Z=10$), 
		where S is chosen as the $d_{xy}$-wave superconductor. 
		We use $\gamma = 1.0$ in all cases. 
		 }
		\label{fig:2Ddwave2barrior}
	\end{figure}

In contrast, $\sigma(eV)$ shows a drastic feature due to the presence of 
an SABS in ARM/$d_{xy}$-waves. 
It is known that the $\sigma(eV)$ in N/$d_{xy}$-waves 
have a ZBCP, 
which is enhanced with increasing $Z$ (see Fig. \ref{fig:2Ddwave2barrior}(a))\cite{TK95}. 
On the other hand, when we consider FM/$d_{xy}$-waves, 
the height of the ZBCP becomes lowered with the increase in 
$h$ as shown in Fig. \ref{fig:2Ddwave2barrior}(b)\cite{Kashiwaya99,Zutic1999,Ting2000}.
In particular, 
when the ferromagnet is fully polarized, 
the ZBCP completely disappears\cite{Kashiwaya99,Zutic1999,Ting2000,Linder2010} 
(see Fig. \ref{fig:2Ddwave2barrior}(b)iii). 
We find that, in ARM/$d_{xy}$-waves, 
the ZBCP appears again due to the presence of $\alpha$ 
(see Fig. \ref{fig:2Ddwave2barrior}(c)). 
Moreover, the height of the ZBCP becomes larger as $\alpha$ increases. 
This $\alpha$ dependence of the $\sigma(eV=0)$ can be understood by the 
spin configuration of the ARM, which is discussed in 
the $s$-wave superconductor junction (see section \ref{swave}). 
As the magnitude of the RSOI increases, the $z$-component of 
spin polarization by the exchange field decreases. 
Additionally, 
the spin polarization by the RSOI does not suppress the tunneling conductance of 
ARM/spin-singlets as mentioned in section \ref{swave}.
This implies
that the ZBCP can be remained in  ARM/$d_{xy}$-waves 
by the RSOI as compared to FPFM/$d_{xy}$-waves (see Fig. \ref{fig:2Ddwave2barrior}(b) and (c)). 

Based on the results in Fig. \ref{fig:2Ddwave2barrior}(c), we discuss 
the physical origin of the presence of ZBCP in an FM/$d_{xy}$-wave with the insulating barrier 
from the aspect of an experiment on 
La$_{0.67}$Sr$_{0.33}$MnO$_{3}$(LSMO)/YBa$_{2}$Cu$_{3}$O$_{7-\delta}$(YBCO) 
with (110) oriented thin film junction\cite{Kashiwaya2004}. 
In the experiment, 
the dependence of the $\sigma(eV)$ on the magnitude of the magnetic field 
applied along in-plane direction has been shown. 
Surprisingly, the ZBCP remains despite of the strongly applied magnetic field, 
where LSMO is known as a half metallic material 
where spin is fully polarized. 
Specifically, the experimental setup \cite{Kashiwaya2004} does not exactly 
correspond with our model. 
In the experiment, the exchange field points the $xy$-plane direction 
while that is parallel to $z$-axis in our model. 
However, also in the junction of the experiment, 
RSOI $\lambda (\hat{{\bm \sigma}}\times{\bm k}) \cdot {\bm i}$ can exist near the interface 
due to the breakdown of the inversion symmetry. 
Here, ${\bm i}$ is a unit vector perpendicular to the interface, $i.e.$, ${\bm i}||{\bm x}$. 
Since this RSOI $\lambda (\hat{{\bm \sigma}}\times{\bm k}) \cdot {\bm x}$ 
induces the $z$-component of the spin-polarization and 
dicreaces the $xy$-plane component of 
the spin polarization induced by the magnetic field and the magnetization, 
LSMO near the interface can behave like the ARM. 
Accordingly, in the light of our theory, 
the ZBCP in the FPFM/$d_{xy}$-wave is allowed in the presence of the RSOI. 
Therefore, the conductance of the Kashiwaya's experiment\cite{Kashiwaya2004} 
may be interpreted from the view point of 
ARM/$d_{xy}$-waves. To compare the experiment and theoretical predoiction 
in detail, it is necessary to take into account 
surface roughness effect.

\subsection{ARM/spin-triplet $p$-wave supercnoductor junction\label{pxandywave}}
In this subsection, we study ARM/spin-triplet $p$-wave superconductor junctions.We mainly focus on  the tunneling conductance for 
ARM/$p_{x}$-waves and ARM/$p_{y}$-waves for several directions of the ${\bm d}$-vector. 
To understand the influence of the RSOI on 
the tunneling conductance, 
we compare the results of ARM/$p_{x}$-waves with ARM/$p_{y}$-waves. 
$g_{{\theta}_{S}}$ in Eq. (\ref{eq:dvectoreqnonhelical}) is given as follows: 
for the $p_{x}$-wave and $p_{y}$-wave symmetries, 
	\begin{eqnarray}
		g_{{\theta}_{S}} =  
		\begin{cases}
			\cos( {\theta}_{S} ) & (p_{x}\text{-wave}) \\
			\sin( {\theta}_{S} ) & (p_{y}\text{-wave})
		\end{cases}. 
	\end{eqnarray}

\begin{figure}[t]
		\centering
		\includegraphics[width=6.4cm]{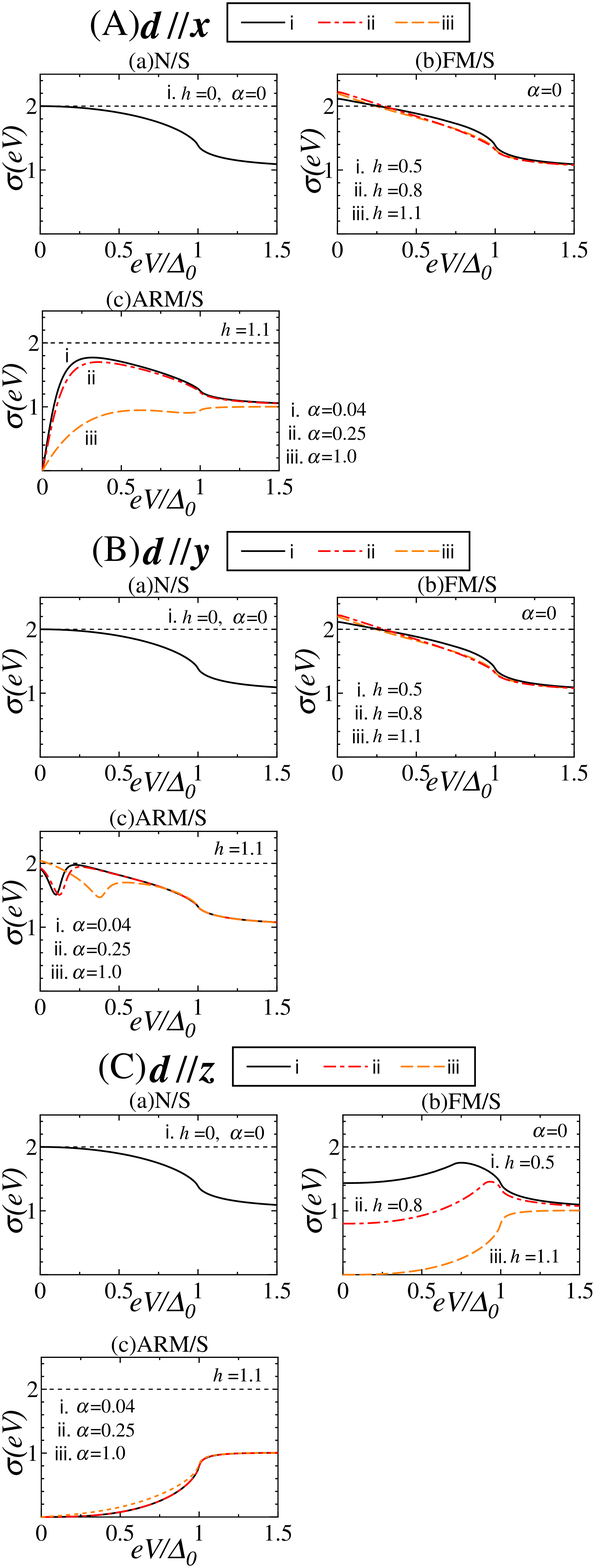}\\
		\caption{(Color online) Normalized $\sigma(eV)$ of two-dimensional
		(a)N/S, (b)FM/S, and (c)ARM/S junctions without insulating barrier ($Z=0$), 
		where S is chosen as the $p_{x}$-wave superconductor 
		for (A)$\bm{d}\parallel\bm{x}$, (B)$\bm{d}\parallel\bm{y}$, and (C)$\bm{d}\parallel\bm{z}$. 
		We use $\gamma = 1$ in all cases. 
		 }
		\label{fig:2Dpxwave}
	\end{figure}

First, Fig. \ref{fig:2Dpxwave} shows the tunneling conductance of 
two-dimensional junctions with $p_{x}$-wave superconductor for $Z=0$.  
Figs. \ref{fig:2Dpxwave} (A), (B), and (C) correspond to  
the cases with $\bm{d}||\bm{x}$, 
$\bm{d}||\bm{y}$, and $\bm{d}||\bm{z}$, respectively. 
The indices (a) , (b), and (c) denote 
N/$p_{x}$-waves, FM/$p_{x}$-waves and ARM/$p_{x}$-waves, respectively. 
For the N/$p_{x}$-waves, 
we have $\sigma(eV=0)=2$ independent of the 
direction of the ${\bm d}$-vector due to the perfect Andreev reflection in $Z=0$\cite{BTK} 
(see Figs. \ref{fig:2Dpxwave}(A)(a), \ref{fig:2Dpxwave}(B)(a), and \ref{fig:2Dpxwave}(C)(a)). 
On the other hand, the ZBC changes from 
$\sigma(eV=0)=2$ in FM/$p_{x}$-waves. 
The change of the $\sigma(eV=0)$ drastically depends on 
the direction of the $ {\bm d}$-vector as well as the magnitude 
of the exchange field $h$. 
When the ${\bm d}$-vector is perpendicular to the exchange 
field, the $\sigma(eV=0)$ changes slightly
(see Figs. \ref{fig:2Dpxwave}(A)(b) and \ref{fig:2Dpxwave}(B)(b)). 
However, when the ${\bm d}$-vector is parallel to the 
exchange field, 
the $\sigma(eV=0)$ is significantly reduced with the magnitude of $h$
(see Fig. \ref{fig:2Dpxwave}(C)(b)). 
Now, let us show $\sigma(eV)$ in ARM/$p_{x}$-waves. 
We find that $\sigma(eV=0)$ is zero for 
${\bm d}||{\bm x}$ and ${\bm d}||{\bm z}$, 
but the $\sigma(eV=0)$ is nonzero only for ${\bm d}||{\bm y}$. 
As we can see from Figs. \ref{fig:2Dpxwave}(A)(c) and (C)(c), 
the inner gap conductance for ${\bm d}||{\bm z}$ is 
insensitive to the magnitude of the RSOI, 
while that for ${\bm d}||{\bm x}$ changes with the magnitude of the RSOI. 
Besides, the $\sigma(eV=0)$ does not strongly depend on the magnitude of the RSOI, 
and $\sigma(eV=0) \sim  2$ is satisfied for ${\bm d}||{\bm y}$ (see Fig. \ref{fig:2Dpxwave}(B)(c)). 
As we will show later, the dependance of $\sigma(eV)$ 
of ARM/$p_{x}$-waves on the direction 
of ${\bm d}$-vector can be explained 
by a winding number, 
which is a topological invariant ensuring the presence of a zero energy SABS. 

\begin{figure}[t]
		\centering
		\includegraphics[width=6.3cm]{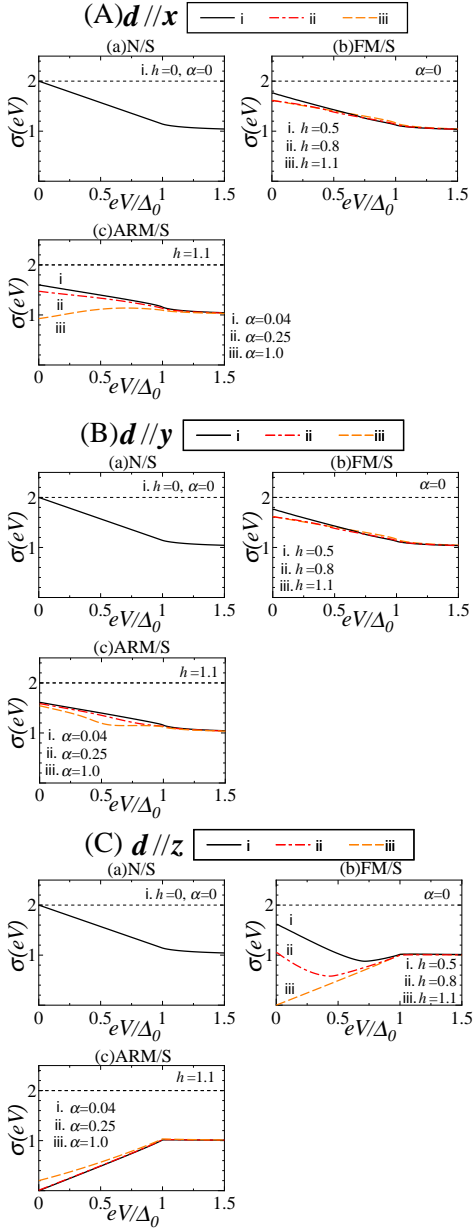}\\
		\caption{(Color online) Normalized $\sigma(eV)$ of two-dimensional
		(a)N/S,(b)FM/S, and (c)ARM/S junctions without insulating barrier ($Z=0$), 
		where SC is chosen as the $p_{y}$-wave superconductor 
		for (A)$\bm{d}\parallel\bm{x}$, (B)$\bm{d}\parallel\bm{y}$, and (C)$\bm{d}\parallel\bm{z}$. 
		We use $\gamma = 1$ in all cases. 
		 }
		\label{fig:2Dpywave}
	\end{figure}

To explain the above anomalous property of $\sigma(eV)$ 
in ARM/$p_{x}$-waves, 
we show the tunneling conductance in two-dimensional ARM/$p_{y}$-waves for $Z=0$. 
Comparing the results in ARM/$p_{y}$-waves with 
those in the ARM/$p_{x}$-waves is important 
because the SABS is absent in junctions with $p_{y}$-wave superconductor 
unlike those with $p_{x}$-wave superconductor\cite{STYY11}. 
Figure \ref{fig:2Dpywave} shows $\sigma(eV)$ of the junctions with $p_{y}$-wave superconductor. 
The indices of Fig. \ref{fig:2Dpywave} 
(A), (B), (C), (a), (b), and (c) 
are the same as those of Fig. \ref{fig:2Dpxwave}, respectively. 
As we can see from  Figs. \ref{fig:2Dpywave}(A)(a), (A)(b), (B)(a), 
(B)(b), (C)(a), and (C)(b) \cite{YTK98,Hirai2003}, 
the behaviors of the inner gap conductance $\sigma(|eV|<{\Delta}_{0})$ in N/$p_{y}$-waves and FM/$p_{y}$-waves 
are qualitatively similar to those in the junctions with $p_{x}$-wave superconductor. 
However, the behaviors of $\sigma(eV=0)$ of the ARM/$p_{y}$-waves 
are qualitatively different from those of the ARM/$p_{x}$-waves. 
In the ARM/$p_{y}$-waves, regardless of the direction of the ${\bm d}$-vector, 
the $\sigma(eV=0)$ is not zero despite of finite $\alpha$. 
To be specific, the $\sigma(eV=0)$ recovers as $\alpha$ increases 
for $\bm{d}||\bm{z}$ (see Fig. \ref{fig:2Dpywave}(C)(c)), 
while that for $\bm{d}||\bm{x}$ and $\bm{d}||\bm{y}$ is slightly reduced 
(see Figs. \ref{fig:2Dpywave}(A)(c) and \ref{fig:2Dpywave}(B)(c)). 
As we describe below, these behaviors of the $\sigma(eV=0)$ of 
ARM/$p_{y}$-waves can be understood by the spin 
configuration of the ARM. 
In spin-triplet superconductor junctions for $\bm{d}||\bm{z}$, 
when an electron with up-spin injects, 
the Andreev reflected hole has down-spin 
similar to the spin-singlet superconductor junction cases. 
This indicates that the $\sigma(|eV|<{\Delta}_{0})$ for ${\bm d}||{\bm z}$ is 
suppressed by the exchange field 
and is enhanced by the RSOI as shown 
in Figs. \ref{fig:2Dpywave}(C)(b) and \ref{fig:2Dpywave}(C)(c). 
On the other hand, when an electron with up-spin injects, 
the Andreev reflection must occur with an up-spin hole 
for $\bm{d}||\bm{x}$ and $\bm{d}||\bm{y}$. 
Hence, the $\sigma(eV)$ is not reduced in the junctions 
with the FPFM for $\bm{d}||\bm{x}$ and $\bm{d}||\bm{y}$ (see Figs. \ref{fig:2Dpywave}(A)(b)iii and (B)(b)iii). 
However, in the ARM, the RSOI reduces the $z$-component of the spin polarization 
as we discussed in section \ref{swave}. 
Accordingly, the $\sigma(|eV|<{\Delta}_{0})$ decreases 
in the ARM/$p_{y}$-waves with $\bm{d}||\bm{x}$ and $\bm{d}||\bm{y}$ 
on the contrary to those with $\bm{d}||\bm{z}$. 
Therefore, the discussion about the spin configuration supports our calculations. 

Now, we discuss the results of ARM/$p_{x}$-waves. 
In ARM/$p_{x}$-waves, 
the dependance of the $\sigma(eV=0)$ on the direction of ${\bm d}$-vector is 
qualitatively different with that in ARM/$p_{y}$-waves. 
In addition, as we mentioned above, one of the important difference between the superconducting tunnel junctions 
with $p_{x}$-wave superconductor and those with $p_{y}$-wave superconductor 
is whether the SABS can exist or not. 
To understand the behavior of the $\sigma(eV=0)$ 
in ARM/$p_{x}$-waves, 
we introduce a winding number $W$ for the one-dimensional limit ($k_{y}=0$). 
Here, $W$ takes an integer and is defined by a chiral operator 
$\bar{\Gamma}$ \cite{STYY11,Kobayashi} and 
a BdG Hamiltonian $\bar{H}$, 
\begin{eqnarray}
	W \equiv \frac{-1}{4 \pi i} \int d {k_{x}} [ \bar{\Gamma} {\bar{H}}^{-1}(\bm{k}) {\partial}_{k_{x}} {\bar{H}} ({\bm{k}}) ], 
\label{eq:windingnumber}
\end{eqnarray}
where the chiral operator anti-commutes with the BdG Hamiltonian, 
and the line integral in Eq. (\ref{eq:windingnumber}) should be performed in the first Brillouin zone. 
When the winding number $W$ is nonzero, 
the SABS exists at the surface of the superconductor. 
For a spin-triplet superconductor, 
a chiral operator generally depends on the direction of the ${\bm d}$-vector 
(see Appendix \ref{winding}). 
Particularly, for the $p_{x}$-wave superconductor, the chiral operator leading to a nontrivial $W$ is given by 
	\begin{eqnarray}
	   && \bar{\Gamma} = \begin{cases} 
	   		  {\hat{\sigma}}_{z} {\hat{\tau}}_{y} & ({\bm{d}}||{\bm{x}})\\
			  {\hat{\sigma}}_{0} {\hat{\tau}}_{x} & ({\bm{d}}||{\bm{y}})\\
			- {\hat{\sigma}}_{x} {\hat{\tau}}_{y} & ({\bm{d}}||{\bm{z}})
			 \end{cases}, \label{rq:chiraloperator}
	\end{eqnarray}
and the resulting $W$ satisfies $W=2$. 
In the ARM/$p_{x}$-waves, 
the SABS is influenced by the RSOI and the exchange field 
through electrons and holes in the ARM. 
From Eq. (\ref{rq:chiraloperator}), we find that the chiral operator anti-commutes with 
the terms of the RSOI $\lambda k_{x} \hat{{\sigma}}_{y} \hat{{\tau}}_{z}$ 
and the exchange field $H \hat{{\sigma}}_{z} \hat{{\tau}}_{z}$ 
only for ${\bm{d}}||{\bm{y}}$. 
This indicates that the chiral symmetry protecting the SABS 
survives under the RSOI and the exchange field 
only when ${\bm{d}}||{\bm{y}}$. 
Therefore, we can understand the RSOI dependance of the tunneling conductance 
in the ARM/$p_{x}$-waves from the topological point of view. 
The discussion about $W$ and the symmetries are given 
in Appendix \ref{winding}. 

\begin{figure}[t]
		\centering
		\includegraphics[width=8cm]{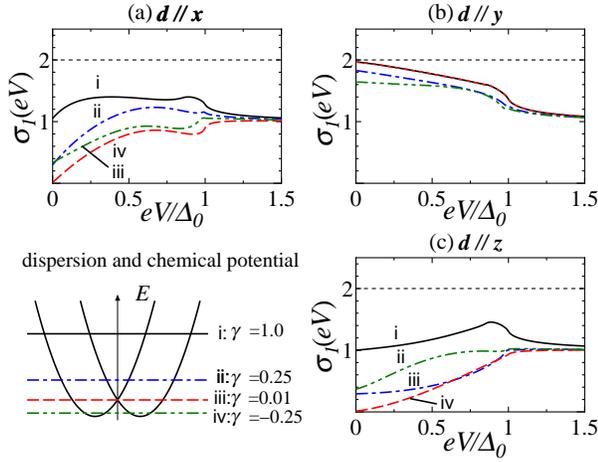}\\
		\caption{(Color online) Normalized tunneling conductance ${\sigma}_{1} (eV)$ of 
		two-dimensional RM/$p_{x}$-wave superconductor junctions for $Z=0$ 
		for various chemical potential ${\mu}_{N}$. 
		${\sigma}_{1} (eV)$ is shown for 
		(a)$\bm{d}||\bm{x}$, (b)$\bm{d}||\bm{y}$, and (c)$\bm{d}||\bm{z}$. 
		We use $\alpha=1$ and $h=0$. 
		}
		\label{fig:2DpxwaveRSOI}
	\end{figure}

\begin{figure}[t]
		\centering
		\includegraphics[width=6.2cm]{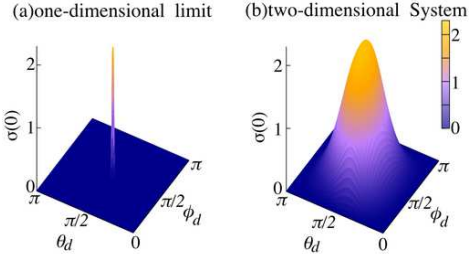}\\
		\caption{(Color online) Normalized zero bias tunneling conductance $\sigma(eV=0)$ 
		of ARM/$p_{x}$-wave superconductor junctions for $Z=0$ as functions of 
		the polar angle ${\theta}_{d}$ and the azimuthal angle ${\phi}_{d}$ of ${\bm d}$-vector (see Fig. \ref{fig:vector}). 
		We use $\gamma = 1.0$, $\alpha = 1.0$, and $h=1.1$ in both 
		(a)one-dimensional limit and (b)two-dimension cases. 
		}
		\label{fig:dvector}
	\end{figure}

Below, with the numerical results, 
we check the validity of the above discussion with $W$. 
As written in Appendix \ref{winding}, 
the RSOI breaks the symmetry protecting the SABS for ${\bm d}||{\bm x}$ and ${\bm d}||{\bm z}$. 
This means that the resulting $W$ is nonzero only for ${\bm d}||{\bm y}$ 
even if the exchange field does not exist. 
Accordingly, the property of the tunneling conductance in ARM/$p_{x}$-waves 
can be realized in the junctions 
with a non-magnetic metal where the RSOI exists, 
which we call a Rashba metal (RM). 
To check whether the property of the tunneling conductance of the ARM/$p_{x}$-waves and 
that of RM/$p_{x}$-waves are similar to each other, 
we calculate the tunneling conductance of the RM/$p_{x}$-waves. 
In Fig. \ref{fig:2DpxwaveRSOI}, the normalized tunneling conductance ${\sigma}_{1}(eV)$, 
where an electron of the outer Fermi surface of RM injects\cite{Yokoyama2006}, 
is shown for the several direction of the ${\bm d}$-vector. 
The details of the formulation is written in Appendix \ref{regionBC}. 
It is found that the ${\sigma}_{1}(eV)$ is suppressed as 
the inner Fermi surface becomes 
smaller for $\bm{d}||\bm{x}$ and $\bm{d}||\bm{z}$ 
(see Figs. \ref{fig:2DpxwaveRSOI}(a) and \ref{fig:2DpxwaveRSOI}(c)). 
Especially, for $\bm{d}||\bm{x}$ and $\bm{d}||\bm{z}$, the ${\sigma}_{1}(eV=0)$ is 
completely reduced for ${\mu}_{N} \rightarrow 0$, 
where the inner Fermi surface of the RM disappears like that of the ARM. 
In contrast, the ${\sigma}_{1}(eV=0)$ is insensitive to $\gamma$ for $\bm{d}||\bm{y}$. 
From these results, it is found that 
the RSOI dominantly contributes to 
the anomalous property of the tunneling conductance in the ARM/$p_{x}$-waves 
while the exchange field does not contribute so much. 
This is consistent with the discussion with the winding number. 
Next, we also calculate how the direction of ${\bm d}$-vector influences on 
$\sigma(eV=0)$ in ARM/$p_{x}$-waves as shown in Fig. \ref{fig:dvector}. 
A sharp peak appears for $\bm{d}||\bm{y}$ in the one-dimensional limit, 
although a broad peak appears in the two-dimensional system. 
The sharp peak in the one-dimensional limit is consistent with our 
discussion based on the winding number. 

\begin{figure}[t]
		\centering
		\includegraphics[width=8cm]{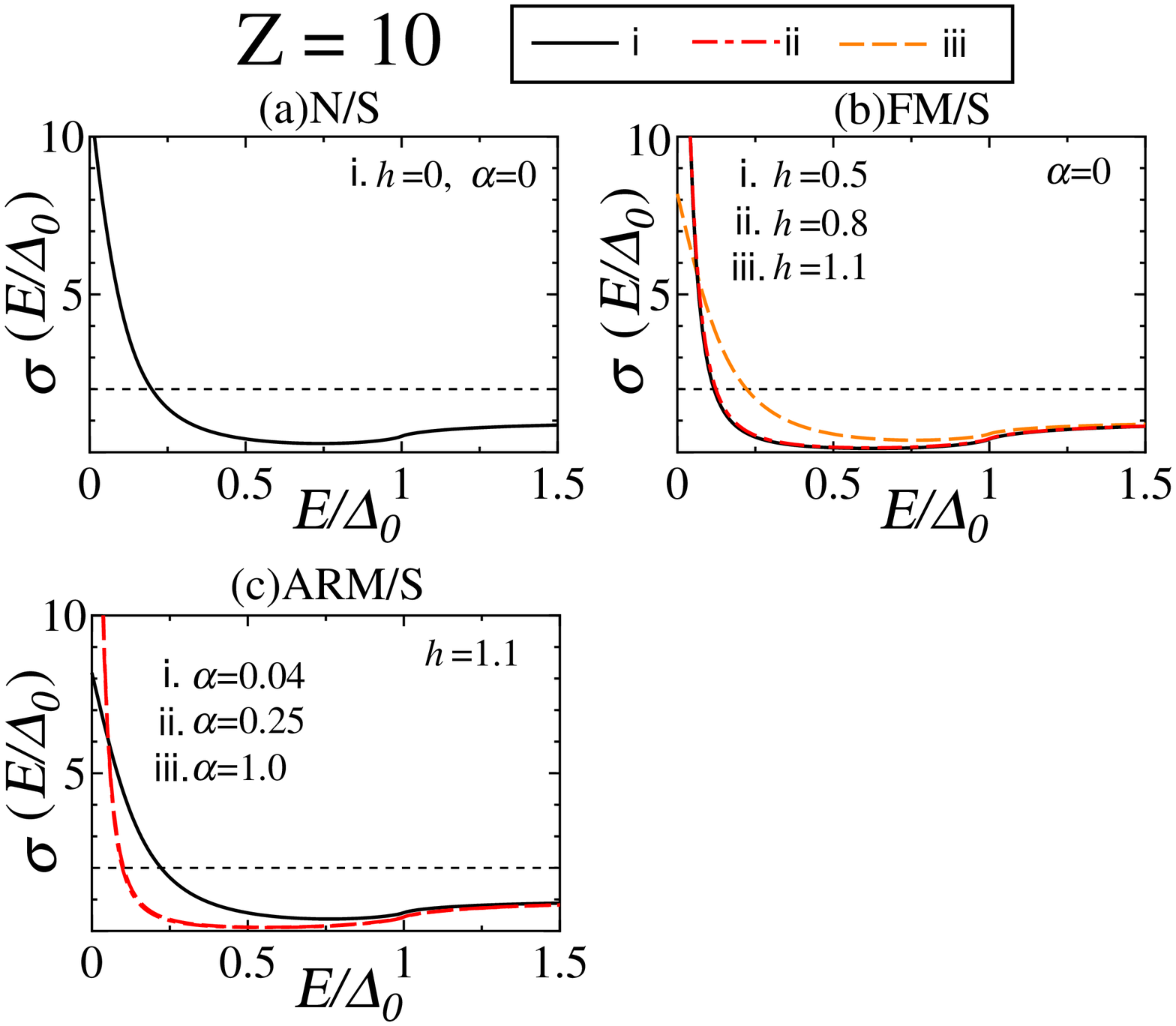}\\
		\caption{(Color online) Normalized $\sigma(eV)$ of two-dimensional
		(a)N/S, (b)FM/S, and (c)ARM/S junctions without insulating barrier ($Z=10$), 
		where S is chosen as the $p_{x}$-wave superconductor for ${\bm d}||{\bm y}$. 
		We use $\gamma = 1.0$ in all cases. 
		 }
		\label{fig:2Dpxwavedybarrior}
	\end{figure}

Finally, $\sigma(eV)$ in ARM/$p_{x}$-waves for high barrier case ($Z=10$) is studied. 
Figure \ref{fig:2Dpxwavedybarrior} shows the obtained $\sigma(eV)$ of 
N/$p_{x}$-waves(a), FM/$p_{x}$-waves(b), and ARM/$p_{x}$-waves(c) for ${\bm d}||{\bm y}$. 
In the N/$p_{x}$-waves, a ZBCP appears (see Fig. \ref{fig:2Dpxwavedybarrior}(a)) 
due to the existence of the SABS regardless of the direction of $\bm{d}$-vector\cite{TK95,kashiwaya00}. 
As we have mentioned already, 
the inner gap conductance does not decrease in FM/spin-triplets 
when $\bm{d}$-vector is perpendicular to the exchange field\cite{Hirai2001,Hirai2003}. 
For this reason, the ZBCP exists for ${\bm d}||{\bm x}$ and ${\bm d}||{\bm y}$ 
(see Fig. \ref{fig:2Dpxwavedybarrior}(b)) 
while the height of the ZBCP is reduced by the exchange field for ${\bm d}||{\bm z}$. 
In ARM/$p_{x}$-waves, 
the $\sigma(eV=0)$ is zero for ${\bm d}||{\bm x}$ and ${\bm d}||{\bm z}$ similarly to the cases for  $Z=0$, and 
the ZBCP appears only when $\bm{d}||\bm{y}$ 
(see Fig. \ref{fig:2Dpxwavedybarrior}(c)). 
In cantrast, the ZBCP does not appear regardless of the change of $\alpha$ and $h$ 
in superconducting tunnel junctions with $p_{y}$-wave superconductor. 
This is natural because the SABS does not 
exist at the surface of $p_{y}$-wave superconductor\cite{kashiwaya00,YTK98}. 
	\begin{table*}
	\centering
		\caption{
		Summary of the behavior of the tunneling conductance $\sigma(eV)$ 
		for the transparent limit and the high barrier case. 
		The first column shows the symmetry of the pair potential in the superconductor. 
		At the first row, X/S and X/I/S indicate the junction 
		for the transparent limit and high barrier case, respectively. 
		Here, this X denotes an N, FM, or ARM. 
		}
		\begin{tabular}{c c c c c c} \hline \hline
			&
			$s$($d_{{x}^{2} - {y}^{2}}$)-wave \ &
			$d_{xy}$-wave \ &
			$p_{x}$-wave (${\bm d}||{\bm x}$) \ &
			$p_{x}$-wave (${\bm d}||{\bm y}$) \ &
			$p_{x}$-wave (${\bm d}||{\bm z}$) \\ \hline 
			\multicolumn{1}{c}{ N/S \ }  &
		 	\multicolumn{1}{c}{ $\sigma(0)=2$ \cite{Andreev,BTK} \ } &
			\multicolumn{1}{c}{ $\sigma(0)=2$ \cite{TK95,kashiwaya00} \ } &
			\multicolumn{1}{c}{ $\sigma(0)=2$ \cite{Sengupta2001} \ } &
			\multicolumn{1}{c}{ $\sigma(0)=2$ \cite{Sengupta2001} \ } &
			\multicolumn{1}{c}{ $\sigma(0)=2$ \cite{Sengupta2001,YTK98,TTKK2002} }  \\ 
			\multicolumn{1}{c}{N/I/S \ } &
			\multicolumn{1}{c}{U(V)-shape \cite{Bardeen1961} \ } &
			\multicolumn{1}{c}{ZBCP \cite{TK95,kashiwaya00} \ } &
			\multicolumn{1}{c}{ZBCP \cite{Sengupta2001} \ } &
			\multicolumn{1}{c}{ZBCP \cite{Sengupta2001} \ } &
			\multicolumn{1}{c}{ZBCP \cite{Sengupta2001,YTK98,TTKK2002} }   \\ \hline
			\multicolumn{1}{c}{FM/S \ } &
			\multicolumn{1}{c}{$\sigma(0) \rightarrow 0$ for $h \rightarrow 1$\cite{Meservey1994, Beenakker95} \ } &
			\multicolumn{1}{c}{$\sigma(0) \rightarrow 0$ for $h \rightarrow 1$\cite{Kashiwaya99,Zutic1999,Zutic2000,Ting2000} \ } &
			\multicolumn{1}{c}{$\sigma(0) \simeq 2$\cite{Hirai2001,Hirai2003} \ } &
			\multicolumn{1}{c}{$\sigma(0) \simeq 2$\cite{Hirai2001,Hirai2003} \ } &
			\multicolumn{1}{c}{$\sigma(0) \rightarrow 0$ for $h\rightarrow 1$\cite{Hirai2001,Hirai2003} } \\ 
			\multicolumn{1}{c}{ FM/I/S \ } &
			\multicolumn{1}{c}{ \ U(V)-shape \cite{Meservey1994,Beenakker95,Linder2010} \ } &
			\multicolumn{1}{c}{ \ No ZBCP for $h\geq 1$\cite{Kashiwaya99,Zutic1999,Zutic2000,Ting2000,Linder2010} \ } &
			\multicolumn{1}{c}{ \ ZBCP\cite{Hirai2001,Hirai2003} \ } & 
			\multicolumn{1}{c}{ \ ZBCP\cite{Hirai2001,Hirai2003} \ } &
			\multicolumn{1}{c}{ \ No ZBCP for $h\geq1$\cite{Hirai2001,Hirai2003} } \\ \hline
			\multicolumn{1}{c}{ ARM/S \ } &
			\multicolumn{1}{c}{ \ $\sigma(0) > 0$ for $\alpha>0$ \ }  &
			\multicolumn{1}{c}{ \ $\sigma(0) > 0$ for $\alpha>0$ \ }  &
			\multicolumn{1}{c}{ \ $\sigma(0) = 0$ for $\alpha>0$ \ }  &
			\multicolumn{1}{c}{ \ $\sigma(0) \simeq 2$ \ }   &
			\multicolumn{1}{c}{ \ $\sigma(0) = 0$ } \\ 
			\multicolumn{1}{c}{ ARM/I/S \ } &
			\multicolumn{1}{c}{ \ U(V)-shape \ } &
			\multicolumn{1}{c}{ \ ZBCP for $\alpha>0$ \ } &
			\multicolumn{1}{c}{ \ No ZBCP \ } &
			\multicolumn{1}{c}{ \ ZBCP \ } &
			\multicolumn{1}{c}{ \ No ZBCP }  \\  \hline 
		\end{tabular}
		\label{table:summary}
	\end{table*}

\section{Relevance to the pairing symmetry in ${Sr}_{2}Ru{O}_{4}$ \label{Sr2RuO4}}
	\begin{figure}[t]
		\centering
		\includegraphics[width=6.2cm]{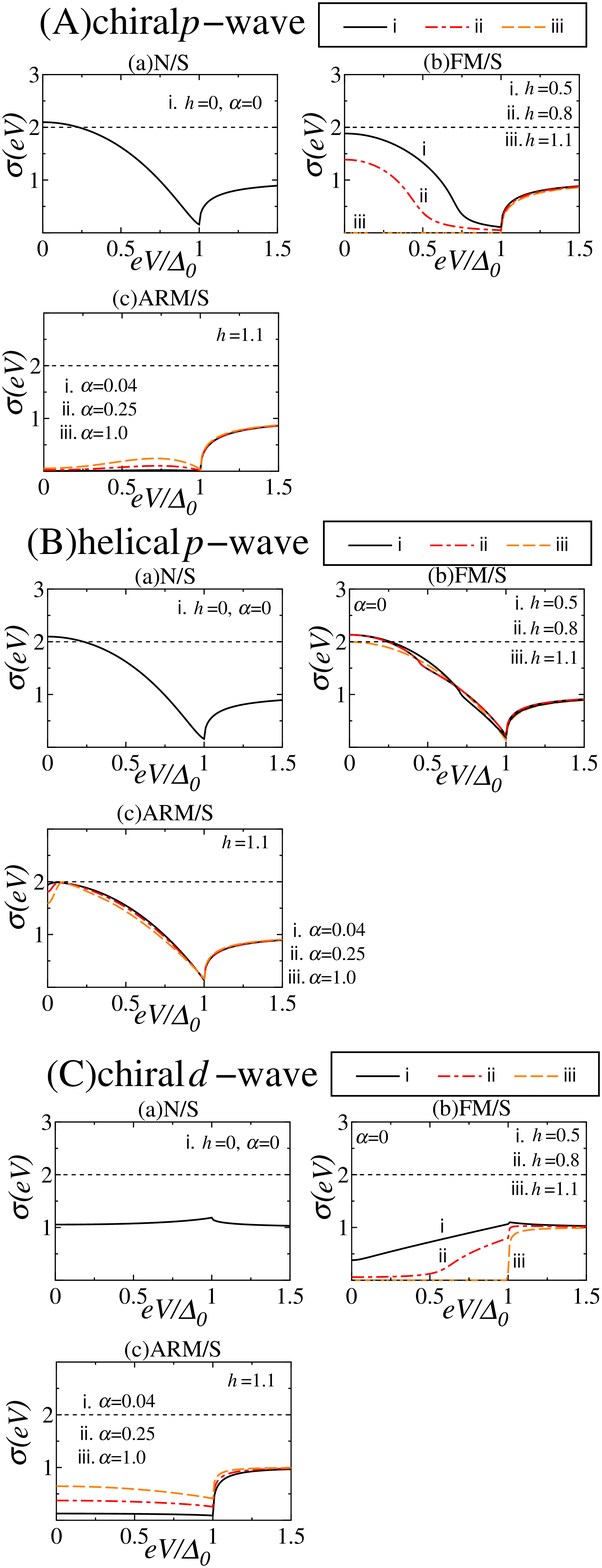}\\
		\caption{(Color online) Normalized $\sigma(eV)$ of 
		(a)N/S, (b)FM/S, and (c)ARM/S junctions  
		where S is chosen as the   
		(A)chiral $p$-wave, (B)helical $p$-wave, and 
		(C)chiral $d$-wave superconductors. 
		We use $Z=5$ and $\gamma = 1.0$ in all cases.
		 }
		\label{fig:SrRuO}
	\end{figure}
In this section, we study the tunneling conductance $\sigma(eV)$ in 
ARM/S junctions where a chiral $p$-wave, helical $p$-wave, and chiral $d$-wave
are chosen as the pairing symmetry in S, respectively. 
Based on the obtained results, 
we suggest a new direction to decide the pairing of ${Sr}_{2}Ru{O}_{4}$. 
To calculate $\sigma(eV)$ in the systems corresponding to experiments of the tunneling spectroscopy, 
we focus on the low transparent junctions with $Z=5$. 

It is noted that chiral $p$-wave pairing is one of 
the promising candidate of the 
pairing symmetry in Sr$_{2}$RuO$_{4}$ \cite{Maeno2012} 
where the $\bm{d}$-vector is along the $z$-axis. 
$g_{{\theta}_{S}}$ is given by 
\begin{equation}
	g_{{\theta}_{S}} = \exp (i {\theta}_{S}) ,
\end{equation}
with $\bm{d}||\bm{z}$. 
First, for an N/chiral $p$-wave, 
the resulting conductance 
has a broad ZBCP reflecting on the linear dispersion of the SABS parallel 
to the interface as shown in Fig. \ref{fig:SrRuO}(A)(a) 
\cite{Kashiwaya11, YTK97, Honerkamp}. 
Then, in FM/chiral $p$-waves, 
the inner gap conductance $\sigma(|eV|<{\Delta}_{0})$  
decreases with the increase in $h$ 
since we consider the cases for ${\bm d}||{\bm z}$
\cite{Hirai2001,Hirai2003} 
(see Fig. \ref{fig:SrRuO}(A)(b)). 
As a limiting case,  
the inner gap conductance is completely suppressed in an FPFM/chiral $p$-wave 
(see Fig. \ref{fig:SrRuO}(A)(b)iii) \cite{Hirai2003,Linder2010}. 
By the comparison with the $\sigma(|eV|<{\Delta}_{0})$ in the FPFM/chiral $p$-wave, 
that in ARM/chiral $p$-waves slightly recovers 
in the presence of the RSOI (see Fig. \ref{fig:SrRuO}(A)(c)). 

Next, we look at the helical $p$-wave case\cite{IHSYMTS07}, where the pair potential 
is given by 
\begin{eqnarray}
	w_{1 {\theta}_{S}} = \cos(\theta_{S}) \ , \ w_{2 {\theta}_{S}} = \sin(\theta_{S}). 
\end{eqnarray}
Time reversal symmetry is not broken in this state. 
There has been a theoretical proposal that 
the helical $p$-wave pairing can be possible by tuning the direction of 
the ${\bm d}$-vector of Sr$_{2}$RuO$_{4}$ \cite{Tada,Ueno}. 
Then, two branches of SABS are generated as a Kramers pair. 
Also in an N/helical $p$-wave, 
the $\sigma(eV)$ has a broad ZBCP
reflecting the linear dispersions of SABS 
crossing zero energy at $k_{y}=0$ similar 
to that in chiral $p$-wave superconductor 
junctions\cite{YTK97} (see Fig. \ref{fig:SrRuO}(B)(a)). 
However, for FM/helical $p$-waves, 
the broad ZBCP remains even for $h>1.0$ 
since, in these cases, the direction of the 
${\bm d}$-vector is in the $xy$-plane \cite{Hirai2001,Hirai2003} 
(see Fig. \ref{fig:SrRuO}(B)(b)). 
On the other hand, for ARM/helical $p$-waves, 
the $\sigma(|eV|<{\Delta}_{0})$ is not seriously suppressed and 
has a small dip around zero-bias voltage \cite{Mukherjee} as shown in 
Fig. \ref{fig:SrRuO}(B)(c). 
In addition, the dip gets bigger as magnitude of the RSOI increases. 
This feature is different from  that in 
ARM/chiral $p$-wave superconductor junction. 

Finally, we calculate $\sigma(eV)$ for chiral $d$-wave junctions 
where the time reversal symmetry is broken 
similar to the case of chiral $p$-wave pairing. 
$f_{{\theta}_{S}}$ is given by 
\begin{equation}
	f_{{\theta}_{S}} = \exp(2i\theta_S). 
\end{equation}
For N/chiral $d$-waves, 
$\sigma(eV)$ 
has almost flat line shape as a function of bias voltage \cite{Kashiwaya2014}. 
Although two branches of the SABS exist, they do not cross $E=0$ at $k_{y}=0$ 
by contrast to chiral $p$-wave and helical $p$-wave pairing cases. 
Then, the contribution from $E=0$ is not large and the resulting 
$\sigma(eV)$ does not have a ZBCP (see Fig. \ref{fig:SrRuO}(C)(a)) \cite{Kashiwaya2014}. 
When we consider FM/chiral $d$-waves, 
the $\sigma(|eV|<{\Delta}_{0})$ is 
reduced with the increase in the magnitude of the exchange field 
since chiral $d$-wave symmetry belongs to a spin-singlet pairing. 
Especially, $\sigma(eV=0)=0$ is satisfied in the junction with the FPFM (see Fig. \ref{fig:SrRuO}(C)(b)iii). 
Similar to the $\sigma(|eV|<{\Delta}_{0})$ in the ARM/spin-singlets shown in the previous section,  
that in ARM/chiral $d$-waves is enhanced with the increase 
of the magnitude of the RSOI (see Fig. \ref{fig:SrRuO}(C)(c)). 

As a summary of the results, if we consider only the N/S junctions, it is difficult to distinguish between 
chiral $p$-wave and helical $p$-wave pairings\cite{Kashiwaya11,YTK97,Kashiwaya2014}. 
For the FM/S junctions with sufficient 
large magnitude of the spin-polarization, 
it is also difficult to distinguish the chiral $d$-wave from 
the chiral $p$-wave\cite{Kashiwaya2014,Hirai2001,Hirai2003}. 
However, for the ARM/S junctions, the qualitative line shapes of $\sigma(eV)$ 
has a different feature for each pairing. 
Therefore, the ARM is useful to classify 
three pairings which have the SABS with linear dispersions. 

\section{Conclusion \label{conclusion}}
In this paper, we have theoretically studied 
tunneling conductance between 
ARM/S 
junctions for various types of the pairing symmetry in S. 
For the ARM/spin-singlet 
superconductor junction, the magnitude of 
the inner gap conductance is enhanced as compared to 
that in the FPFM junction. 
It is  noted that the 
ZBCP recovers in the ARM/$d_{xy}$-wave superconductor junction by the RSOI
while that is completely 
suppressed in the FPFM/$d_{xy}$-wave superconductor junction. 
In a previous work\cite{Kashiwaya2004}, 
the anomalous behavior of the conductance in LSMO/YBCO junctions 
has not been reported, and its origin has not been discovered. 
Our obtained results can explain 
the ZBCP in LSMO/YBCO junctions  in the presence of 
large magnitude of the exchange field. 
Due to the absence of the inversion symmetry, 
RSOI is induced near the interface of LSMO. 
Then, it is natural to speculate that 
LSMO can behave like the ARM near the interface. 
Based on this, the robust ZBCP reported in 
LSMO/YBCO junctions seems to be reasonable \cite{Kashiwaya2004}. 

We have also studied the tunneling conductance in 
the ARM/$p_{x}$-wave superconductor junctions. 
It has been revealed whether the ZBCP remains 
or not critically depends on the direction of the $\bm{d}$-vector in 
ARM/$p_{x}$-wave superconductor junctions, and 
this can be understood by using the winding number $W$. 
In addition, we have calculated 
the tunneling conductance in the ARM/S junction, 
where the symmetry of S is 
the chiral $p$-wave, helical $p$-wave, and chiral $d$-wave pairings. 
We have shown that 
these three types of pairings show qualitatively 
different line shapes of tunneling conductance. 
Our obtained results are useful to determine the pairing symmetry of 
superconductor Sr$_{2}$RuO$_{4}$. 

In this paper, we have focused on the quasiparticle 
tunneling in ARM/S  junctions. 
It is a challenging problem to study Josephson current 
in S/ARM/S junctions since an SABS \cite{tanaka97,kashiwaya00} 
seriously influences on 
the magnitude of Josephson current at low temperatures. 
Although, theoretical study about N/S or S/N/S junctions  
in the presence of RSOI in N has been done 
in some works \cite{TomoYokoyama,Cayao}, 
Josephson current  in S/ARM/S junction has not been revealed 
particularly for unconventional superconductors yet.
We are planning to study this issue near future. 

\section{Acknowledgements}\label{sec_Ack}
This work was supported by 
the Grant-in Aid for Scientific Research on Innovative Areas ``Topological Material Science" 
(Grant No. 15H05853), 
Grant-in-Aid for Scientific Research B (Grant No. 15H03686), 
Grant-in-Aid for Challenging Exploratory Research (Grant No. 15K13498), and 
Grant-in-Aid for JSPS Fellows (Grant No. 13J06466 and No. 13J03141) (K. T  and S. K).  
\appendix
\section{Winding number in $p_x$-wave superconductors \label{winding}}
We here discuss the winding number of $p_x$-wave superconductor of one-dimensional limit, 
which guarantees the existence of a Majorana edge state and complements our numerical results. 
We start from the BdG Hamiltonian of a one-dimensional $p_x$-wave superconductor 
\begin{align}
	{\bar{H}}_{\rm BdG}(k_x) = (2t \cos (k_x) -\mu) {\hat{\sigma}}_0 {\hat{\tau}}_z + \bar{\Delta} (k_x), \label{eq:HBdG}
\end{align}
with
\begin{align}
	\bar{\Delta} (k_x) = \begin{cases} 
		-\Delta_0 \sin k_x {\hat{\sigma}}_z {\hat{\tau}}_x & (\bm{d} \parallel {\bm{x}})\\  
		\Delta_0 \sin k_x {\hat{\sigma}}_0 {\hat{\tau}}_y & (\bm{d} \parallel {\bm{y}})\\ 
		\Delta_0 \sin k_x {\hat{\sigma}}_x {\hat{\tau}}_x & (\bm{d} \parallel {\bm{z}}) 
	\end{cases},
\end{align} 
where $\mu$ is the chemical potential and $\Delta_0$ the amplitude of 
the gap-function. 
${\hat{\sigma}}_i$ $({\hat{\tau}}_i)$ ($i=0,x,y,z$) are 
the identity matrix and the Pauli matrices in the spin (Nambu) space. 
This system satisfies the time-reversal symmetry 
$\bar{T} {\bar{H}}_{BdG}(k_x) {\bar{T}}^{-1} ={\bar{H}}_{BdG}(-k_x)$ and 
the particle-hole symmetry $\bar{C} {\bar{H}}_{BdG}(k_x) {\bar{C}}^{-1} =-{\bar{H}}_{BdG}(-k_x)$ 
by $\bar{T}=i {\hat{\sigma}}_y {\hat{\tau}}_0 K$ and $\bar{C}= {\hat{\sigma}}_0 {\hat{\tau}}_x \bar{K}$, 
where $\bar{K}$ is the complex conjugation. 

If the BdG Hamiltonian has a chiral operator $\bar{\Gamma}$; $i.e.$, $\{ \bar{\Gamma}, {\bar{H}}_{BdG}(k_x)\} =0$, 
then the winding number is defined by\cite{Wen:2002,Beri:2010,STYY11,Schnyder:2011}
\begin{align}
	W \equiv \frac{-1}{4 \pi i} \int_{-\pi}^{\pi} d k_x \; \trace[\bar{\Gamma} {\bar{H}}_{\rm BdG}(k_x)^{-1} \partial_{k_x} {\bar{H}}_{\rm BdG}(k_x)], \label{eq:winding}
\end{align}
which takes an integer. 
In time-reversal invariant superconductors, 
the combination of time-reversal operator $\bar{T}$ and particle-hole operator $\bar{C}$ becomes 
the chiral operator ${\bar{\Gamma}}_0= -i \bar{C} \bar{T}$. 
Due to the inversion symmetry, we notice that whereas Eq. (\ref{eq:winding}) with ${\bar{\Gamma}}_0$ yields 
a nontrivial winding number in spin-singlet superconductors, 
it leads to $W=0$ in spin-triplet superconductors\cite{Kobayashi}. 
Thus, in order to pursue a nontrivial winding number in a spin-triplet pairing, 
we require an aid of material dependent symmetries in addition to $\bar{T}$ and $\bar{C}$. 

Equstion (\ref{eq:HBdG}) possesses the spin-rotational symmetries: 
${\bar{U}}_x = i {\hat{\sigma}}_x {\hat{\tau}}_z$, ${\bar{U}}_y = -i {\hat{\sigma}}_y {\hat{\tau}}_0$, and ${\bar{U}}_z = i {\hat{\sigma}}_z {\hat{\tau}}_z$, 
which satisfy $[{\bar{U}}_i, {\bar{H}}_{\rm BdG} (k_x)] =0$ when the $\bm{d}$-vector is parallel to the $i$ direction. 
Taking into account this additional symmetry, 
we can define a spin dependent chiral operator ${\bar{\Gamma}}_i \equiv  \bar{C} \bar{T} {\bar{U}}_i$, and 
Eq. (\ref{eq:winding}) 
with ${\bar{\Gamma}}_i$ leads to, for each direction of the $\bm{d}$-vector,
\begin{align}
	W = \begin{cases} 2 & 0 < \mu < 2 t \\ 0 & \text{otherwise}\end{cases},  \label{eq:W}
\end{align} 
where $W=2$ indicates the presence of Majorana Kramer's pair at both ends. 
 
On the other hand, in our numerical result, we found that the zero-bias conductance peak is suppressed 
when $\bm{d} \parallel \bm{x}$ and $\bm{d} \parallel \bm{z}$. 
To explain this suppression from Eq. (\ref{eq:W}), 
we consider how the Rashba spin-orbit interaction (RSOI) and 
the exchange field affect the Majorana Kramer's pair by adding the terms 
\begin{align}
	\bar{H}' = \lambda \sin k_x {\hat{\sigma}}_y {\hat{\tau}}_z + H {\hat{\sigma}}_z {\hat{\tau}}_z
\end{align}
into the BdG Hamiltonian, 
where the parameters $\lambda$ and $H$ indicate the amplitude of RSOI and the exchange field, respectively. 
We readily find that the first term breaks the spin-rotational 
symmetries ${\bar{U}}_x$ and ${\bar{U}}_z$, $i.e.$, 
the winding number survives only when $\bm{d} \parallel {\bm{y}}$. 
In addition, although a Majorana Kramer's pair is fragile against the exchange effect, 
we have the effective time-reversal symmetry $\bar{T}'= \bar{T} {\bar{U}}_y$ for the $y$-direction, 
which keeps the Majorana Kramer's pair intact even when the Zeeman effect is 
present\cite{Dumitrescu:2013,Dumitrescu:2014}. 
As a result, the topological argument is consistent with our calculation of the tunneling conductance. 

%
%
%
\section{Formulation for the tunneling conductance in Rashba metal with exchange field/superconductor junctions 
\label{regionBC}}
\begin{figure}[t]
	\centering
	\includegraphics[width=5.5cm]{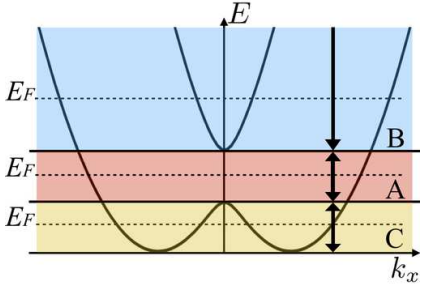}\\
	\caption{(Color online) Energy spectrum of the Rashba metal with exchange field. There are three regions A, B, and C, depending on $\mu$.}
	\label{fig:phase}
\end{figure}
\begin{figure}[t]
	\centering
	\includegraphics[width=5.5cm]{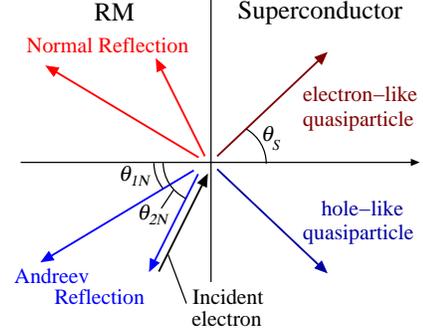}\\
	\caption{(Color online) Schematic illustration of the scattering process. ${\theta}_{1(2)N}$ is an incident angle of an
		electron with momentum 
		$k_{1(2)}$ respect to the interface normal. 
		${\theta}_{S}$ denotes the direction of motions of quasiparticles in S measured from the interface normal.}
	\label{fig:scattering2}
\end{figure}
We here show formulations for the tunneling conductance of 
Rashba metal (RM)/insulator/superconductor junctions 
in the presence of the exchange 
field in the RM where the number of the Fermi surfaces is two. 
The BdG Hamiltonian of this system is already given by 
Eqs. (\ref{eq:BdG}),  (\ref{eq:NormalHamiltonian}), and (\ref{eq:PairPotential}). 
In this appendix, we shift its attention from the ARM ($\left| \mu \right|<H$) to the RM ($\left| \mu \right|>H$) 
and derive the wave functions for $x<0$ and the tunneling conductance. 
Therefore, the wave functions for $x<0$ and 
the tunneling conductance are mainly introduced in this section. 
If we choose $H=0$, the resulting tunneling conductance corresponds with that shown 
in section \ref{pxandywave} (see Fig. \ref{fig:2DpxwaveRSOI}). 

The dispersion in the RM with the exchange field is shown in Fig. \ref{fig:phase}. 
From Fig. \ref{fig:phase}, it is found that we can define three regions (A, B, and C) depending on 
the parameters: $H$, $\lambda$, and ${\mu}_{N}$. 
The ARM is realized in the region A with $\left| \mu \right|<H$. 
On the other hand, 
the RM is realized in the region B (C) with 
$\mu>H$ ($- \frac{{(m \lambda)}^{2} + H^{2}}{2 {(m \lambda)}^{2}}<\mu<-H$). 
In the regions B and C, 
we have two Fermi surface unlike in the region A. 
Thus, we need to take into account the inner Fermi surface in addition to the outer Fermi surface. 
Interestingly, 
the inner and the outer Fermi surfaces have different helicity each other in the region B 
while they have the same spin helicity in the region C. 
The scattering process for the regions B and C is shown in Fig. \ref{fig:scattering2}. 
${\theta}_{1N}$ (${\theta}_{2N}$) is an incident angle of the momentum 
for the outer (inner) Fermi surface $k_{1}$ ($k_{2}$). 
The momenta correspond to what we have shown in section \ref{formalism}: 
\begin{eqnarray}
	k_{1} &=& 
	\sqrt{2m \biggl( {\mu}_{N} + m{\lambda}^{2} + \sqrt{ {(m {\lambda}^{2})}^{2} + 2 m {\lambda}^{2} {\mu}_{N} +H^2 } \biggr) }, \nonumber \\
	k_{2} &=& 
	\sqrt{2m \biggl( {\mu}_{N} + m{\lambda}^{2} - \sqrt{ {(m {\lambda}^{2})}^{2} + 2 m {\lambda}^{2} {\mu}_{N} +H^2 } \biggr) }, \nonumber
\end{eqnarray}
and the $y$-component of all momenta is given by 
\begin{eqnarray}
	k_{y} = k_{1} \sin {\theta}_{1N} = k_{2} \sin {\theta}_{2N} = k_{S} \sin {\theta}_{S} .
\end{eqnarray}
In what follows, we discuss the formulations in the region B (i) and C (ii). 

(i) In this paragraph, 
we show the formulation for the tunneling conductance of 
a two-dimensional RM with the exchange field (the region B)/insulator/superconductor junction. 
The wave functions are represented by using the eigenfunctions of the BdG Hamiltonian for $\mu>H$. 
First, we introduce the wave function in the case where an electron of the 
outer Fermi surface injects, 
\begin{widetext}
\begin{eqnarray}
	\psi (x<0,y) &=& \frac{1}{\sqrt{2}}  e^{i k_{y} y} \Biggl(  
	                  e^{i k_{1} \cos {\theta}_{1N} x}           \begin{bmatrix}            s        \\ 1 \\           0         \\  0  \end{bmatrix}
	       + r_{1(2)} e^{-i k_{1} \cos {\theta}_{1N} x}  \begin{bmatrix}      {s}^{\ast}  \\ 1 \\           0         \\  0  \end{bmatrix}
	       +   a_{1(2)} e^{ i k_{1} \cos {\theta}_{1N} x}  \begin{bmatrix}             0       \\ 0 \\   - {s}^{\ast}    \\  1 \end{bmatrix} \nonumber \\
	       &+& r_{2(1)} e^{-i K_{eBx} x}  \begin{bmatrix}      t_{B1e}    \\ 1 \\          0          \\  0  \end{bmatrix}
	       +   a_{2(1)} e^{ i K_{hBx} x}  \begin{bmatrix}             0       \\ 0 \\  - t_{B1h}      \\  1 \end{bmatrix}
 					    \Biggr),  
\end{eqnarray}
\end{widetext}
\begin{eqnarray}
	&&\ \ \ \ \ \ \ \ \ \ \ \ \ \ \ \ \ \ \ \ \ \ \ \ s    = -\frac{i \lambda k_{1} e^{-i{\theta}_{1N}}}{{\xi}_{{\bm k}_{1}}  + H} , \nonumber \\
	&&\ \ \ \ \ \ \ \ \ \ \ \ \ \ \ \ \ {t}_{B1e(h)} = - \frac{\lambda ( - i K_{e(h)Bx} + k_{y}) }{\xi_{{\bm k}_{2}} + H} , \nonumber \\ 
	&&K_{e(h)Bx} = \begin{cases} 
				k_{2} \cos {\theta}_{2N} & (k_{1} \sin {\theta}_{1N} < k_{2}) \\
				+(-)i \sqrt{{k_{1}}^{2} {\sin}^{2} {\theta}_{1N} - {k_{2}}^{2}} & (k_{1} \sin {\theta}_{1N} > k_{2})  \end{cases}. \nonumber
\end{eqnarray}
For $k_{1} \sin {\theta}_{1N} > k_{2}$, the normal and Andreev reflections from 
inner Fermi surface become the evanescent waves. 
Next, we introduce the wave function in the case where an electron of the inner Fermi surface injects, 
\begin{widetext}
\begin{eqnarray}
	 \psi (x<0,y) &=& \frac{1}{\sqrt{2}} e^{i k_{y} y} \Biggl(  
	                  e^{i k_{2} \cos {\theta}_{2N} x}  \begin{bmatrix}         t_{B2}      \\ 1 \\           0             \\  0  \end{bmatrix}
	        + r_{1} e^{-i k_{1} \cos {\theta}_{1N} x}  \begin{bmatrix}     {s}^{\ast}     \\ 1 \\           0             \\  0  \end{bmatrix} 
	        + a_{1} e^{ i k_{1} \cos {\theta}_{1N} x}  \begin{bmatrix}            0          \\ 0 \\     - {s}^{\ast}      \\  1 \end{bmatrix} \nonumber \\
	     &+& r_{2} e^{-i k_{2} \cos {\theta}_{2N} x}  \begin{bmatrix} {t_{B2}}^{\ast} \\ 1 \\            0            \\  0  \end{bmatrix}
	        + a_{2} e^{ i k_{2} \cos {\theta}_{2N} x}  \begin{bmatrix}             0         \\ 0 \\ - {t_{B2}}^{\ast} \\  1 \end{bmatrix}
 					    \Biggr) , 
\end{eqnarray}
\end{widetext}
\begin{eqnarray}
	&&
	    s    = -\frac{i \lambda k_{1} e^{-i{\theta}_{1N}}}{{\xi}_{{\bm k}_{1}}  + H} , \nonumber \\
	&&
	{t}_{B2} = -\frac{i \lambda k_{2} e^{-i{\theta}_{2N}}}{\xi_{{\bm k}_{2}} + H}. \nonumber 
\end{eqnarray}
We assume that the wave function in the junction satisfies the boundary condition given by Eq. (\ref{eq:boundary}).
The obtained tunneling conductance is given as follows: 
\begin{eqnarray}
   	\sigma (E) &=& \frac{1}{2} {\sigma}_{1}(E) + \frac{1}{2} {\sigma}_{2}(E) , \\
	{\sigma}_{1}(E) &=& \frac{\int {\sigma}_{1S} (E,{\theta}_{S}) d {\theta}_{S}}{\int {\sigma}_{1N} (E,{\theta}_{S}) d {\theta}_{S}} , \label{eq:APPCC11} \\
	{\sigma}_{2}(E) &=& \frac{\int {\sigma}_{2S} (E,{\theta}_{S}) d {\theta}_{S}}{\int {\sigma}_{2N} (E,{\theta}_{S}) d {\theta}_{S}} . \label{eq:APPCC12}
\end{eqnarray}
Here, ${\sigma}_{1(2)}(E)$ means normalized tunneling conductance 
when an electron from the outer (inner) Fermi surface injects. 
In addition, ${\sigma}_{i S}$ (${\sigma}_{i N}$) represents tunneling conductance between 
the ARM/S (the ARM/normal metal (${\Delta}_{0}=0$)) junction, 
where $i=1,2$. 
In Eqs. (\ref{eq:APPCC11}) and (\ref{eq:APPCC12}), 
${\sigma}_{1S}(E,{\theta}_{S})$ and ${\sigma}_{2S}(E,{\theta}_{S})$ are given by 
\begin{widetext}
\begin{eqnarray}
	{\sigma}_{1S}(E,{\theta}_{S}) &=& \begin{cases}
							4e \biggl( (1 + {|a_{1}|}^{2} - {|r_{1}|}^{2}) \bigl( \frac{{k}_{1} \cos {\theta}_{1N}}{m} ({|s|}^{2} + 1 ) - i \lambda ( s - {s}^{\ast}) \bigr) \\
						\ \ \ \ \ \ \ \ +({|a_{2}|}^{2} - {|r_{2}|}^{2}) \bigl(  \frac{{k}_{2} \cos {\theta}_{2N}}{m} ({|{t}_{B1e}|}^{2} + 1 ) + i \lambda ( {t}_{B1e} - {{t}_{B1e}}^{\ast}) \bigr) \biggr) 
									& (k_{1} \sin {\theta}_{1N} < k_{2}) \\
							4e           (1 + {|a_{1}|}^{2} - {|r_{1}|}^{2}) \bigl( \frac{{k}_{1} \cos {\theta}_{1N}}{m} ({|s|}^{2} + 1 ) - i \lambda ( s - {s}^{\ast}) \bigr) 
									& (k_{1} \sin {\theta}_{1N} > k_{2}) \end{cases} , \\
	{\sigma}_{2S}(E,{\theta}_{S}) &=& 	4e \biggl( (1 + {|a_{2}|}^{2} - {|r_{2}|}^{2}) 
			\bigl( \frac{{k}_{2} \cos {\theta}_{2N}}{m} ({|{t}_{B2}|}^{2} + 1 ) - 
			i \lambda ( {t}_{B2} - {{t}_{B2}}^{\ast}) \bigr) \nonumber \\
			&&\ \ \ \ \ \ \ \ + ({|a_{1}|}^{2} - {|r_{1}|}^{2}) \bigl(  \frac{{k}_{1} \cos {\theta}_{1N}}{m} ({|s|}^{2} + 1 ) - 
			i \lambda ( s - {s}^{\ast}) \bigr) \biggr) .
\end{eqnarray}
\end{widetext}

(ii) Next, 
we show the formulation for the tunneling conductance of 
a two-dimensional RM with the exchange field (the region C)/insulator/superconductor junction.
Wave functions are represented by using the eigenfunctions of the BdG Hamiltonian 
for $- \frac{{(m \lambda)}^{2} + H^{2}}{2 {(m \lambda)}^{2}}<\mu<-H$. 
First, we introduce the wave function in the case where an electron of 
the outer Fermi surface injects, 
\begin{widetext}
\begin{eqnarray}
	\psi (x,y) &=& \frac{1}{\sqrt{2}}  e^{i k_{y} y} \Biggl(  
	                  e^{i k_{1} \cos {\theta}_{1N} x}       \begin{bmatrix}            s        \\ 1 \\           0         \\  0  \end{bmatrix} 
	       + r_{1} e^{-i k_{1} \cos {\theta}_{1N} x}  \begin{bmatrix}      {s}^{\ast}  \\ 1 \\           0         \\  0  \end{bmatrix}
	       +   a_{1} e^{ i k_{1} \cos {\theta}_{1N} x}  \begin{bmatrix}             0       \\ 0 \\  - {s}^{\ast}    \\  1 \end{bmatrix} \nonumber \\
	     &+& r_{2} e^{-i K_{2ex} x}  \begin{bmatrix}      t_{C1e}    \\ 1 \\          0          \\  0  \end{bmatrix}
	       +   a_{2} e^{ i K_{2hx} x}  \begin{bmatrix}             0       \\ 0 \\  - t_{C1h}      \\  1 \end{bmatrix}
 					    \Biggr),  
\end{eqnarray}
\end{widetext}
\begin{eqnarray}
&& \ \ \ \ \ \ \ \ \ \ \ \ \ \ \ \ \ \ \ \ \ 
	      s = -\frac{i \lambda k_{1} e^{-i{\theta}_{1N}}}{{\xi}_{{\bm k}_{1}}  + H} ,\nonumber \\
&& \ \ \ \ \ \ \ \ \ \ \ \ \ \ \ \ \ 
	{t}_{C1e(h)} = - \frac{\lambda ( i K_{e(h)Cx} + k_{y}) }{\xi_{{\bm k}_{2}} + H} , \nonumber \\ 
&&
	K_{2e(h)x} = \begin{cases} 
				k_{2} \cos {\theta}_{2N} & (k_{1} \sin {\theta}_{1N} < k_{2}) \\
				+(-)i \sqrt{{k_{1}}^{2} {\sin}^{2} {\theta}_{1N} - {k_{2}}^{2}} & (k_{1} \sin {\theta}_{1N} > k_{2})  \end{cases}. \nonumber 
\end{eqnarray}
For $k_{1} \sin {\theta}_{1N} > k_{2}$, the normal and Andreev reflections 
from inner Fermi surface become the evanescent waves. 
Next, we introduce the wave function in the case where an electron of the inner Fermi surface injects, 
\begin{widetext}
\begin{eqnarray}
	\psi (x,y) &=& \frac{1}{\sqrt{2}}  e^{i k_{y} y} \Biggl(  
	                  e^{i k_{2} \cos {\theta}_{2N} x}    \begin{bmatrix}    {t_{C2}}^{\ast}      \\ 1 \\           0             \\  0  \end{bmatrix}
	       + r_{1} e^{-i k_{1} \cos {\theta}_{1N} x}  \begin{bmatrix}     {s}^{\ast}     \\ 1 \\           0             \\  0  \end{bmatrix}
	       +   a_{1} e^{ i k_{1} \cos {\theta}_{1N} x}  \begin{bmatrix}            0          \\ 0 \\    - {s}^{\ast}      \\  1 \end{bmatrix} \nonumber \\
	     &+& r_{2} e^{-i k_{2} \cos {\theta}_{2N} x}  \begin{bmatrix}         {t_{C2}}     \\ 1 \\            0            \\  0  \end{bmatrix}
	       +   a_{2} e^{ i k_{2} \cos {\theta}_{2N} x}  \begin{bmatrix}             0         \\ 0 \\      - {t_{C2}}      \\  1 \end{bmatrix}
 					    \Biggr) , 
\end{eqnarray}
\end{widetext}
\begin{eqnarray}
&&\ \ \ \ \ \ \ \ \ \ \ \ \ \ \
	    s    = -\frac{i \lambda k_{1} e^{-i{\theta}_{1N}}}{{\xi}_{{\bm k}_{1}}  + H} ,\nonumber \\
&&\ \ \ \ \ \ \ \ \ \ \ \ \ \ \ 
	{t}_{C2} = -\frac{i \lambda k_{2} e^{-i{\theta}_{2N}}}{\xi_{{\bm k}_{2}} + H} .\nonumber 
\end{eqnarray}
We assume that the wave function satisfies the boundary condition given by Eq.(\ref{eq:boundary}).
The obtained tunneling conductance is given as follows: 
\begin{eqnarray}
   	\sigma (E) &=& \frac{1}{2} {\sigma}_{1}(E) + \frac{1}{2} {\sigma}_{2}(E) , \\
	{\sigma}_{1}(E) &=& \frac{\int {\sigma}_{1S} (E,{\theta}_{S}) d {\theta}_{S}}{\int {\sigma}_{1N} (E,{\theta}_{S}) d {\theta}_{S}} , \label{eq:APPCC21} \\
	{\sigma}_{2}(E) &=& \frac{\int {\sigma}_{2S} (E,{\theta}_{S}) d {\theta}_{S}}{\int {\sigma}_{2N} (E,{\theta}_{S}) d {\theta}_{S}} . \label{eq:APPCC22}
\end{eqnarray}
Here, ${\sigma}_{1(2)}(E)$ means normalized tunneling conductance 
when an electron from the outer (inner) Fermi surface injects. 
In addition, ${\sigma}_{i S}$ (${\sigma}_{i N}$) represents tunneling conductance between 
the ARM/S (the ARM/normal metal (${\Delta}_{0}=0$)) junction, 
where $i=1,2$. 
In Eqs. (\ref{eq:APPCC21}) and (\ref{eq:APPCC22}), 
${\sigma}_{1S}(E,{\theta}_{S})$ and ${\sigma}_{2S}(E,{\theta}_{S})$ are given by 
\begin{widetext}
\begin{eqnarray}
	{\sigma}_{1S}(E,{\theta}_{S}) &=& \begin{cases}
							4e \biggl( (1 + {|a_{1}|}^{2} - {|r_{1}|}^{2})                   \bigl( \frac{{k}_{1} \cos {\theta}_{1N}}{m} ({|s|}^{2} + 1 ) - i \lambda ( s - {s}^{\ast}) \bigr) \\
						\ \ \ \ \ \ \ \ + ({|a_{2}|}^{2} - {|r_{2}|}^{2}) \bigl(  \frac{{k}_{2} \cos {\theta}_{2N}}{m} ({|{t}_{C1e}|}^{2} + 1 ) - i \lambda ( {t}_{C1e} - {{t}_{C1e}}^{\ast}) \bigr) \biggr) 
									& (k_{1} \sin {\theta}_{1N} < k_{2}) \\
							4e           (1 + {|a_{1}|}^{2} - {|r_{1}|}^{2})                                           \bigl( \frac{{k}_{1} \cos {\theta}_{1N}}{m} ({|s|}^{2} + 1 ) - i \lambda ( s - {s}^{\ast}) \bigr) 
									& (k_{1} \sin {\theta}_{1N} > k_{2}) \end{cases} , \nonumber \\ \\
	{\sigma}_{2S}(E,{\theta}_{S}) &=& 	4e \biggl( (1 + {|a_{2}|}^{2} - {|r_{2}|}^{2})) 
		\bigl( \frac{{k}_{2} \cos {\theta}_{2N}}{m} ({|{t}_{C2}|}^{2} + 1 ) - 
		i \lambda ( {t}_{C2} - {{t}_{C2}}^{\ast}) \bigr) \nonumber \\
		&& \ \ \ \ \ \ \ \ + ({|a_{1}|}^{2} - {|r_{1}|}^{2})) \bigl(  \frac{{k}_{1} \cos {\theta}_{1N}}{m} ({|s|}^{2} + 1 ) - 
		i \lambda ( s - {s}^{\ast}) \bigr) \biggr). 
\end{eqnarray}
\end{widetext}

\bibliography{ARMtunnel}

\end{document}